\documentclass[12pt, a4paper]{article}
\usepackage{cite}
\usepackage{amsmath,amssymb}
\input{colordvi.tex}
\usepackage{comment}
\usepackage{bm}
\usepackage{url}
\usepackage{ifpdf}
\ifpdf        
  \usepackage{graphicx, hyperref, xcolor}     %   usepackage without driver option
\else     % For (p)LaTeX + dvipdfmx
  \usepackage[dvipdfmx]{graphicx, hyperref, xcolor}     %   usepackage with driver option
 \fi

\definecolor{rossoferrari}{HTML}{D9073D}
\definecolor{mediumblue}{HTML}{0000CD}
\hypersetup{% hyperref option list
setpagesize=false,
bookmarksnumbered=true,%
bookmarksopen=true,%
colorlinks=true,%
linkcolor=rossoferrari,
urlcolor=mediumblue,
citecolor=mediumblue,
}

\begin{document}

%%%%%%%%%%%%%%%%%%%%%%%%%%%%%%%%%%%%%%%%%%%%%%%%%%
\begin{titlepage}

\begin{center}

\hfill UT-19-18\\

\vskip .75in

{\Large \bf 
Vector Coherent Oscillation Dark Matter
}

\vskip .75in

{\large
Kazunori Nakayama$^{(a,b)}$
}

\vskip 0.25in

$^{(a)}${\em Department of Physics, Faculty of Science,\\
The University of Tokyo,  Bunkyo-ku, Tokyo 113-0033, Japan}\\[.3em]
$^{(b)}${\em Kavli IPMU (WPI), The University of Tokyo,  Kashiwa, Chiba 277-8583, Japan}

\end{center}
\vskip .5in

\begin{abstract}

We construct a model of hidden massive vector boson dark matter as a homogeneous coherent oscillation in the entire universe without any dangerous instability. We make use of a particular form of the vector boson coupling to a scalar field through the gauge kinetic function. This scenario may be distinguished from other dark matter models through the observation of statistical anisotropy in the dark matter isocurvature perturbation.\footnote{
		{\bf [Note added in the latest arXiv version (Mar.2023)]} This scenario turned out to be inconsistent with cosmological observations taking account of the statistical anisotropy of the curvature fluctuation~\cite{Nakayama:2020rka}. Besides this issue, there have been several mistakes and misleading points, so I modified some texts and added footnotes in Sec.\,3.4 and Sec.\,4 to the old version for readers' convenience.}

\end{abstract}

\end{titlepage}

%\tableofcontents

\renewcommand{\thepage}{\arabic{page}}
\setcounter{page}{1}
\renewcommand{\thefootnote}{\#\arabic{footnote}}
\setcounter{footnote}{0}
%%%%%%%%%%%%%%%%%%%%%%%%%%%%%%%%%%%%%%%%%%%%%%%%%%

\newpage

\tableofcontents

%%%%%%%%%%%%%%%%%%%%%%%%%%%%%%%%%%%%%%%%%%%%%%%%%%
\section{Introduction}
\label{sec:Intro}
%%%%%%%%%%%%%%%%%%%%%%%%%%%%%%%%%%%%%%%%%%%%%%%%%%

Recently there are increasing interests on the ultra light dark matter (DM) scenario such as axion-like particle or hidden vector boson~\cite{Jaeckel:2010ni,Arias:2012az}.
These light DM particles are assumed to have extremely tiny coupling to the standard model sector, but still many ideas to detect or constrain such a scenario have been proposed.

One of the issues of the ultra light DM scenario is to obtain a correct relic abundance of DM.
For the case of scalar field, it is well known that the scalar field with mass much smaller than the Hubble scale during inflation can develop a homogeneous condensate during inflation and it exhibits a coherent oscillation at some later epoch: the so-called misalignment mechanism.
This coherent oscillation acts as a non-relativistic matter and its abundance can fit the observed value of DM density in the present universe.
A famous example is the QCD axion~\cite{Preskill:1982cy,Abbott:1982af,Dine:1982ah,Kim:1986ax,Kawasaki:2013ae}.

The case of hidden vector boson is more involved, since a free massive vector boson cannot have a homogenous condensate during inflation, as briefly reviewed in the next section.
Instead, people often consider other production mechanisms of hidden vector boson DM: production through the $h(\phi) F_{\mu\nu} \widetilde F^{\mu\nu}$ coupling~\cite{Agrawal:2018vin,Co:2018lka,Bastero-Gil:2018uel} where $h(\phi)$ is some function of a scalar field $\phi$, $F_{\mu\nu}$ and $\widetilde F^{\mu\nu}$ are the field strength of the vector boson and its dual, respectively, production from the $h(\phi) F_{\mu\nu} F^{\mu\nu}$ coupling~\cite{Dror:2018pdh} and also purely gravitational production during and at the end of inflation~\cite{Graham:2015rva,Ema:2019yrd}. They can yield right amount of vector DM, some of which make ultra-light vector boson possible DM candidate, although none of these are related with the ``coherent oscillation'' of vector boson.\footnote{
	In this paper, only the oscillation of the homogenous condensate in the entire universe is called as coherent oscillation.
}

The vector coherent oscillation as (ultra-light) DM was considered in Refs.~\cite{Arias:2012az,AlonsoAlvarez:2019cgw} although their is a flaw in their mechanism.
In Refs.~\cite{Arias:2012az,AlonsoAlvarez:2019cgw} the vector boson $A_\mu$ coupling to the Ricci curvature $R$ is introduced in the Lagrangian as $\mathcal L \sim (1/12)R A_\mu A^\mu$ to cancel the effective Hubble mass term and make vector boson effectively massless.
However, such a coupling necessarily induces a ghost instability for the longitudinal vector mode~\cite{Dvali:2007ks,Himmetoglu:2008zp,Himmetoglu:2009qi,Karciauskas:2010as}.
The roles of $RA_\mu A^\mu$ coupling have been considered in the context of magnetogenesis~\cite{Turner:1987bw} and the vector curvaton~\cite{Dimopoulos:2006ms,Dimopoulos:2008rf,Dimopoulos:2008yv}, although all of these attempts eventually suffer from the serious ghost instability.

The main purpose of this paper is to make a consistent model for a vector coherent oscillation.
To this end, we borrow the idea of the vector curvaton scenario utilizing the kinetic coupling of the form $f^2(\phi) F_{\mu\nu} F^{\mu\nu}$~\cite{Dimopoulos:2007zb,Dimopoulos:2009am,Dimopoulos:2009vu,Dimopoulos:2010xq} with some scalar field $\phi$, which does not suffer from serious instability.
As explained in detail in the main text, by assuming some specific time dependence of $f(\phi)$, it is possible to effectively cancel the Hubble mass term during inflation so that the vector field develops a homogeneous condensate during inflation while the theory reduces to just a free massive vector field at later epoch. 
In such a setup the vector field begins a homogeneous coherent oscillation well after inflation and it behaves as non-relativistic DM.
The vanishing Hubble mass during inflation also implies the development of long-wave fluctuation and the function $f(\phi)$ controls the property of the isocurvature perturbation of the vector DM.

In Sec.~\ref{sec:rev} we briefly review why a free massive vector boson cannot develop a homogeneous condensate during inflation and a problem of introducing the $RA_\mu A^\mu$ coupling.
In Sec.~\ref{sec:co} we give a detailed study of the model with $f^2(\phi) F_{\mu\nu} F^{\mu\nu}$ coupling. After discussing the dynamics of the homogeneous condensate, we give analysis of the fluctuation. In particular, the statistically anisotropic DM isocurvature perturbation can be a unique signal of this vector DM scenario.
Sec.~\ref{sec:dis} is devoted to conclusions and discussion.

%%%%%%%%%%%%%%%%%%%%%%%%%%%%%%%%%%%%%%%%%%%%%%%%%%
\section{Free massive vector field and its extension}
\label{sec:rev}
%%%%%%%%%%%%%%%%%%%%%%%%%%%%%%%%%%%%%%%%%%%%%%%%%%

%%%%%%%%%%%%%%%%%%%%%%%%%%%%%%%%%%%%%%%%%%%%%%%%%%
\subsection{Action}
%%%%%%%%%%%%%%%%%%%%%%%%%%%%%%%%%%%%%%%%%%%%%%%%%%

The action of the massive vector field $\mathcal A_M$ is given by
\begin{align}
	S = \int d^4x \sqrt{-g}\left( -\frac{1}{4}g^{MN}g^{KL}\mathcal F_{MK}\mathcal F_{NL} - \frac{1}{2}m^2 g^{MN} \mathcal A_M \mathcal A_N \right),
\end{align}
where $\mathcal F_{MN} = \partial _M \mathcal A_N - \partial_N \mathcal A_M$,\footnote{
	Note that $\mathcal F^{MN} = g^{MK} g^{NL}\mathcal F_{KL} \neq \partial^M \mathcal A^N-\partial^N \mathcal A^M$ for a general metric.
} and we assume the Friedmann-Robertson-Walker (FRW) metric:
\begin{align}
	g_{MN}={\rm diag}(-1,a^2(t),a^2(t),a^2(t)),
\end{align}
with $a(t)$ being the cosmic scale factor. It is often convenient to use the conformal time $d\tau = dt /a$ and rewrite the action as
\begin{align}
	S =  \int d\tau d^3x \left( -\frac{1}{4}\eta^{\mu\rho}\eta^{\nu\sigma}F_{\mu\nu}F_{\rho\sigma} - \frac{a^2}{2}m^2\eta^{\mu\nu}A_\mu A_\nu \right),
\end{align}
where $\eta^{\mu\nu} = {\rm diag}(-1,1,1,1)$ and we defined $A_{\mu}= (A_0,A_i)\equiv (a\mathcal A_0, \mathcal A_i)$ and $F_{\mu\nu}=\partial_\mu A_\nu-\partial_\nu A_\mu$. In what follows, tensors with greek superscript are understood as those raised by $\eta^{\mu\nu}$.
This form makes the conformal nature of the vector boson clear: the metric dependence completely disappears in the massless limit $m\to 0$.
The vector boson mass term arises either by the Higgs mechanism or the Stuckelberg mechanism. In the former case, as far as the radial component of the Higgs field is heavy enough, all the phenomenology is indistinguishable from the latter one.

%The variation gives
%%
%\begin{align}
%	&\delta F_{\mu\nu} = (\delta_{\mu\alpha}\delta_{\nu\beta} - \delta_{\nu\alpha}\delta_{\mu\beta})  \delta(\partial_\alpha A_\beta),\\
%	&\delta (\eta^{\mu\rho}\eta^{\nu\sigma}F_{\mu\nu}F_{\rho\sigma}) = 4 F^{\alpha\beta} \delta(\partial_\alpha A_\beta).
%\end{align}
%%
The equation of motion is
\begin{align}
	\partial_\mu F^{\mu\nu} - a^2m^2 A^\nu = 0.
\end{align}
Here $F^{\mu\nu} = \eta^{\mu\rho}\eta^{\nu\sigma} F_{\rho\sigma}$ and $A^\nu = \eta^{\nu\mu}A_\mu$.
It gives
\begin{align}
	&\partial_i A_i'-\nabla^2 A_0 + a^2m^2 A_0 = 0,  \label{constraint} \\
	&A_i'' - \nabla^2 A_i + a^2m^2 A_i + \partial_i( -A_0' + \partial_i A_i )= 0.
\end{align}
where $' = \partial /\partial \tau = a \partial /\partial t$. There is no gauge invariance for the massive vector boson, so we cannot take a gauge to make equations simpler.

In Fourier space, defining
\begin{align}
	A_\mu(\vec x,\tau) = \int \frac{d^3k}{(2\pi)^3} A_\mu(\vec k,\tau) e^{i\vec k \cdot \vec x},
\end{align}
with $A_\mu(\vec k,\tau)=A_\mu^*(-\vec k,\tau)$, one can explicitly solve the constraint equation for $A_0$ (\ref{constraint}) as\footnote{
	For notational simplicity we use the same character for the vector field in the position space $A_\mu(\tau,\vec x)$ and momentum space $A_\mu(\vec k,\tau)$. We believe that which one is used is clear in the context and this does not lead to any confusion.
}
\begin{align}
	A_0(\vec k,\tau) = \frac{-i\vec k\cdot {\vec A'}}{k^2+a^2m^2} = \frac{-i k A_L'}{k^2+a^2m^2}.
\end{align}
Here the vector boson is decomposed into the transverse and longitudinal ones $\vec A = \vec A_T + \hat k A_L$ where $\hat k \equiv \vec k/k$ the transverse mode satisfies $\vec k\cdot \vec A_T=0$.
The action for the transverse and longitudinal mode is given as
\begin{align}
	& S=S_T + S_L,\\
	&S_T = \int \frac{d^3k d\tau}{(2\pi)^3} \frac{1}{2}\left(  |\vec A_T'|^2 - (k^2+a^2m^2)|\vec A_T|^2 \right),\\
	&S_L=  \int \frac{d^3k d\tau}{(2\pi)^3} \frac{1}{2}\left(  \frac{a^2m^2}{k^2+a^2m^2}|A_L'|^2 - a^2m^2 |A_L|^2 \right).  \label{SL}
\end{align}
%%

%%%%%%%%%%%%%%%%%%%%%%%%%%%%%%%%%%%%%%%%%%%%%%%%%%
\subsection{Zero mode dynamics}
%%%%%%%%%%%%%%%%%%%%%%%%%%%%%%%%%%%%%%%%%%%%%%%%%%

For spatially homogeneous case $A_\mu(t,\vec x) = A_\mu(t)$, the first equation just gives $A_0=0$.
The second equation gives
\begin{align}
	A_i'' + a^2m^2 A_i = 0 ~~~\leftrightarrow~~~ \ddot A_i + H \dot A_i + m^2 A_i=0,
\end{align}
where the dot denotes the derivative with respect to $t$ and $H\equiv \dot a/a$ is the Hubble parameter.
Later we also use the conformal Hubble parameter $\mathcal H \equiv a'/a = aH$.
This has an oscillating solution like $A_i \propto a^{-1/2}(t) \cos(mt)$ for $m\gg H$.
For $H\ll m$, it clearly has a solution $A_i \simeq {\rm const}$. However, it does not mean a homogeneous condensate can be formed during inflation, since the energy density decreases as $\rho \propto a^{-2}$ even if $A_i = {\rm const}$ (see Sec.~\ref{sec:energy}).
It is related to the fact that $A_i$ should be regarded as a ``comoving'' field rather than a physical field, as seen below.

Here it may be worth mentioning that $A_i$ is regarded as a ``comoving field'' while $\overline{A_i} \equiv A_i/a$ as a ``physical'' field.
One can understand this terminology by looking at the kinetic (time derivative) term in the action as
\begin{align}
	S \supset \int d\tau d^3x\,\frac{1}{2}A_i'^2 
	 = \int dt d^3X \frac{1}{2a^2} \dot A_i^2 = \int dt d^3X \left(\frac{1}{2} \dot{\overline{A_i}}^2 -\frac{1}{2}(H^2-\dot H)\overline{A_i}^2\right),
\end{align}
where we have defined physical coordinate $d\vec X \equiv ad\vec x$.
Thus $A_i$ is canonical in the comoving coordinate $(\tau,\vec x)$ while $\overline{A_i}$ is canonical in the physical coordinate $(t,\vec X)$.

In terms of the ``physical'' field $\overline{A_i}$, the equation of motion is expressed as
\begin{align}
	\ddot {\overline {A_i}} + 3H\dot {\overline {A_i}}  + (m^2 + 2H^2+\dot H) \overline{A_i} = 0.  
\end{align}
Note that the Ricci curvature is expressed as $R=6(2H^2+\dot H)$. Thus it exhibits a similar equation as a scalar field. For $m\gg H$, it has an oscillating solution like $\overline{A_i} \propto a^{-3/2}(t) \cos(mt)$. For $m\ll H$, due to the mass term of $2H^2$ one cannot have a solution like $\overline{A_i}\sim {\rm const.}$ Even if we start with some finite value of $\overline{A_i}$ during or before inflation, it is exponentially damped during inflation and we end up with practically vanishing $\overline{A_i}$ after inflation. This is the reason why we cannot have a homogeneous vector condensate in a theory of simple massive vector field.\footnote{
	Ref.~\cite{Nelson:2011sf} missed the Hubble mass term in the equation of motion and incorrectly derived coherent oscillation of a free massive vector field.
}

%%%%%%%%%%%%%%%%%%%%%%%%%%%%%%%%%%%%%%%%%%%%%%%%%%
\subsection{Extension and instability}
%%%%%%%%%%%%%%%%%%%%%%%%%%%%%%%%%%%%%%%%%%%%%%%%%%

The observation in the previous subsection may lead us to introduce a curvature coupling
\begin{align}
	\mathcal L = \frac{1}{2}\xi R \,g^{MN} A_M A_N,
\end{align}
in the Lagrangian with $\xi$ being a constant. The action and equation of motion are the same after one reinterprets $m^2 \to m^2 - \xi R$.
The equation of motion of the zero mode becomes
\begin{align}
	\ddot {\overline {A_i}} + 3H\dot {\overline {A_i}}  + \left(m^2 + \left(\frac{1}{6}-\xi\right)R \right) \overline{A_i} = 0. 
\end{align}
For $\xi=1/6$ one obtains a vanishing Hubble-induced mass term and the equation becomes the same as a minimal massive scalar field. Thus one may have a solution like $\overline{A_i}={\rm const}$ during inflation and hence the vector coherent oscillation occurs at some epoch after inflation.
This curvature coupling of the vector field was considered in the context of magnetogenesis~\cite{Turner:1987bw}, vector curvaton~\cite{Dimopoulos:2006ms,Dimopoulos:2008rf,Dimopoulos:2008yv} and hidden photon DM as a coherent oscillation~\cite{Arias:2012az,AlonsoAlvarez:2019cgw}.

However, this scenario suffers from the ghost instability~\cite{Dvali:2007ks,Himmetoglu:2008zp,Himmetoglu:2009qi,Karciauskas:2010as}.
It is easy to see that the kinetic term of the longitudinal mode (\ref{SL}) has a wrong sign for some finite $k$ after the replacement $m^2 \to m^2 - \xi R$ since one must have $m^2- \xi R < 0$ in order to have a solution like $\overline{A_i}={\rm const}$. In the Higgs picture, the appearance of ghost instability may be understood as an existence of the Higgs kinetic term with a wrong sign in order to obtain a tachyonic mass for the vector boson.
Although small $k$ $(\ll aH)$ modes do not have ghost instability, they must be well inside the horizon $(k\gg aH)$ as time goes back, hence they originate from the ghost regime. It could be healthy if the time scale of the ghost instability is much longer than the Hubble time scale, although it is improbable~\cite{Carroll:2003st}. Thus it is safe to say that we cannot discuss the vector field dynamics in a healthy way.
The situation is more or less the same for other extensions to modify the ``potential'' of the vector field in order to cancel the Hubble mass term somehow~\cite{Ford:1989me}.

%%%%%%%%%%%%%%%%%%%%%%%%%%%%%%%%%%%%%%%%%%%%%%%%%%
\section{Massive vector field with kinetic function}
\label{sec:co}
%%%%%%%%%%%%%%%%%%%%%%%%%%%%%%%%%%%%%%%%%%%%%%%%%%

%%%%%%%%%%%%%%%%%%%%%%%%%%%
\subsection{Action}
%%%%%%%%%%%%%%%%%%%%%%%%%%%

Let us consider the action of massive vector field with kinetic function $f(\phi)$ which is dependent on some scalar field $\phi$:
\begin{align}
	S &= \int d^4x \sqrt{-g} \left(-\frac{1}{4}f^2(\phi) g^{MN} g^{KL} \mathcal F_{MK}\mathcal F_{NL} - \frac{1}{2} m^2 g^{MN} \mathcal A_M \mathcal A_N \right) \\
	& =  \int d\tau d^3x \left( -\frac{1}{4}\eta^{\mu\rho}\eta^{\nu\sigma}f^2(\phi)F_{\mu\nu}F_{\rho\sigma} - \frac{a^2}{2}m^2\eta^{\mu\nu}A_\mu A_\nu \right)
\end{align}
The equation of motion reads
\begin{align}
	\partial_\mu \left( f^2 F^{\mu\nu}\right) - a^2m^2 A^\nu = 0,
\end{align}
which give
\begin{align}
	&f^2(\partial_i A_i'-\nabla^2 A_0) -\partial_i f^2(\partial_i A_0- A_i') + a^2m^2 A_0 = 0,  \label{constraint_f} \\
	&f^2(A_i'' - \nabla^2 A_i) + a^2m^2 A_i + f^2\partial_i( -A_0' + \partial_i A_i ) +f^{2\prime}(A_i'-\partial_i A_0)-\partial_j f^2(\partial_j A_i-\partial_i A_j)= 0.
\end{align}

Repeating the same procedure as the previous section, we obtain the action in the Fourier space for the transverse and longitudinal mode as follows:
\begin{align}
	&S = S_T + S_L,\\
	&S_T = \int \frac{d\tau d^3k}{(2\pi)^3} \frac{1}{2}
	\left( f^2|\vec{A}'_T(k)|^2 -(f^2k^2+a^2m^2) |\vec{A}_T(k)|^2 \right),\\
	&S_L= 
	\int \frac{d\tau d^3k}{(2\pi)^3} \frac{1}{2}
	\left( \frac{f^2 a^2m^2}{f^2k^2+a^2m^2}|{A}'_L(k)|^2 - a^2m^2 |{A}_L(k)|^2 \right).
\end{align}
Since the kinetic term always has a correct sign, there is no ghost instability in this model.
One can further rewrite this action in terms of the canonical field $\vec{A}^c_T(k)\equiv f\vec{A}_T(k)$
and ${A}^c_L(k)\equiv g{A}_L(k)$ where $g\equiv f\sqrt{a^2m^2/(f^2k^2+a^2m^2)}$:
\begin{align}
	&S_T = \int \frac{d\tau d^3k}{(2\pi)^3} \frac{1}{2}
	\left[ |\vec{A}^{c \prime}_T(k)|^2 - \left(k^2+\frac{a^2m^2}{f^2}- \frac{f''}{f} \right) |\vec{A}^c_T(k)|^2 \right], \label{ST_f} \\
	&S_L= 
	\int \frac{d\tau d^3k}{(2\pi)^3} \frac{1}{2}
	\left[|{A}^{c\prime}_L(k)|^2 -\left(k^2 + \frac{a^2m^2}{f^2} - \frac{g''}{g} \right) |{A}^c_L(k)|^2 \right].  \label{SL_f}
\end{align}
Explicit calculation shows
\begin{align}
	%frac{g''}{g}=-\frac{3 a^2m^2 f^2k^2}{(f^2k^2+a^2m^2)^2}\left(\mathcal H-\frac{f'}{f}\right)^2+ \frac{1}{f^2k^2+a^2m^2}\left(f^2k^2\frac{a''}{a}+a^2m^2\frac{f''}{f}\right).
	\frac{g''}{g}=\frac{f''}{f} +  \frac{f^2k^2}{f^2k^2+a^2m^2}\left(
		\frac{a''}{a}-\frac{f''}{f} -\frac{3 a^2m^2}{f^2k^2+a^2m^2}\left(\mathcal H-\frac{f'}{f}\right)^2
	 \right).
\end{align}
Thus the difference between transverse and longitudinal mode appears in the second term of this expression.

%%%%%%%%%%%%%%%%%%%%%%%%%%%%%%%%%%%%%%%%%%%%%%%%%%
\subsection{Energy momentum tensor}  \label{sec:energy}
%%%%%%%%%%%%%%%%%%%%%%%%%%%%%%%%%%%%%%%%%%%%%%%%%%

Before going into the analysis of the dynamics, we give en expression for the energy momentum tensor for later use.
The energy momentum tensor is defined as
\begin{align}
	T_{MN} &= \frac{-2}{\sqrt{-g}}\frac{\delta(\sqrt{-g}\mathcal L)}{\delta g^{MN}} \\
	&= f^2g^{AB}\mathcal F_{MA}\mathcal F_{NB} -\frac{f^2}{4}g_{MN} g^{AB}g^{CD} \mathcal F_{AC} \mathcal F_{BD} + m^2\left( \mathcal A_M \mathcal A_N -\frac{1}{2}g_{MN}g^{AB}\mathcal A_A \mathcal A_B \right).
\end{align}
For the spatially homogeneous case $A_\mu(t,\vec x) = A_\mu(t)$ ($A_0=0$ in this case), we have
\begin{align}
	\rho_A = T_{00} = \frac{1}{2a^2} \left( f^2\dot A_i^2 + m^2 A_i^2 \right)
	=\frac{1}{2}\left[\dot{\overline{A_i}} + \left(H-\frac{\dot f}{f}\right) \overline{A_i}\right]^2 + \frac{m^2}{2f^2} \overline{A_i}^2,
\end{align}
where we have defined a ``physical'' field $\overline{A_i} \equiv f A_i/a$.
Let us take a coordinate such that $A_i=(0,0,A_z)$ without loss of generality. Then we have $T_{0i}=0$ and $T_{ij}=0$ for $i\neq j$. For diagonal components, we obtain
\begin{align}
	T_{xx}=T_{yy}=\frac{f^2}{2}\dot A_z^2-\frac{1}{2}m^2 A_z^2,~~~~~~T_{zz}=-\frac{f^2}{2}\dot A_z^2+\frac{1}{2}m^2 A_z^2.
\end{align}
In the deep oscillating regime $m/f \gg H$ and $f\simeq {\rm const}$, we have $T_{xx}=T_{yy}=T_{zz}=0$ after the oscillation average, implying the zero pressure. Thus the coherently oscillating vector field just behaves as non-relativistic matter and does not induce anisotropic expansion. On the other hand, if the vector condensate during its slow-roll phase is a dominant component of the universe, it induces anisotropic expansion. In our scenario studied below, the vector boson begins a coherent oscillation well before it dominates the universe, and hence the isotropy of the universe is not affected by the vector background.

By using the equation of motion (\ref{eom_f_zero}), we obtain the energy conservation law as
\begin{align}
	\dot \rho_{A} =\dot \rho_A^{(K)} + \dot \rho_A^{(V)} = -\left( 4H + \frac{2\dot f}{f}\right) \rho_A^{(K)} -2H \rho_A^{(V)},
	\label{rhoK}
\end{align}
where
\begin{align}
	\rho_A^{(K)} = \frac{1}{2}\left[\dot{\overline{A_i}} + \left(H-\frac{\dot f}{f}\right) \overline{A_i}\right]^2
	= \frac{1}{2}\left[\dot{\overline{A_i}} + \left(1-\frac{\alpha}{2}\right) H\overline{A_i}\right]^2,~~~~~~
	\rho_A^{(V)} = \frac{m^2}{2f^2} \overline{A_i}^2,
\end{align}
where in the second expression of $\rho_A^{(K)}$ we have substituted $f^2 \propto a^{\alpha}$.
In the deep oscillation regime, we have $f={\rm const}$ and $\rho_A^{(K)} \simeq \rho_A^{(V)}$, hence $\dot\rho_A=-3H \rho_A$, as it should be.

Including the fluctuation, the energy density is expressed as
\begin{align}
	&\rho_A = \rho_{A_T} + \rho_{A_L},\\
	&\rho_{A_T} = \int \frac{d^3k}{(2\pi)^3} \frac{1}{2a^4}
	\left( f^2|\vec{A}'_T(k)|^2 +(f^2k^2+a^2m^2) |\vec{A}_T(k)|^2 \right),\\
	&\rho_{A_L} = 
	\int \frac{d^3k}{(2\pi)^3} \frac{1}{2a^4}
	\left( \frac{f^2 a^2m^2}{f^2k^2+a^2m^2}|{A}'_L(k)|^2 + a^2m^2 |{A}_L(k)|^2 \right).
\end{align}
%%

%%%%%%%%%%%%%%%%%%%%%%%%%%%
\subsection{Zero mode dynamics}  \label{sec:zero}
%%%%%%%%%%%%%%%%%%%%%%%%%%%

First we study the evolution of the zero mode. Assuming $A_\mu(t,\vec x) = A_\mu(t)$ and $\phi(t,\vec x)=\phi(t)$, we immediately find that $A_0=0$ from the constraint equation (\ref{constraint_f}). 
Defining the physical field $\overline{A_i} \equiv f A_i/a$, we find 
\begin{align}
	\ddot {\overline {A_i}} + 3H\dot {\overline {A_i}}  + \left(\frac{m^2}{f^2} + 2H^2+\dot H -H \frac{\dot f}{f} -\frac{\ddot f}{f} \right ) \overline{A_i} = 0
	\label{eom_f_zero}
\end{align}
It obtains an additional effective mass term from the time dependence of the kinetic function $f$. Now we assume $f^2 \propto a^{\alpha}(t)$ during inflation and finally it approaches to $f\to 1$ around the end of inflation. Later we will show an example to realize this scaling by the scalar dynamics. Then the equation of motion becomes
\begin{align}
	\ddot {\overline {A_i}} + 3H\dot {\overline {A_i}}  + \left(\frac{m^2}{f^2} -\frac{(\alpha+4)(\alpha-2)}{4}H^2 + \frac{2-\alpha}{2}\dot H \right ) \overline{A_i} = 0.
\end{align}
Thus choosing $\alpha=2$ or $\alpha=-4$ results in the vanishing Hubble induced mass term and renders $\overline {A_i}$ effectively massless during inflation, assuming $m/f \ll H$ always holds during inflation~\cite{Dimopoulos:2009am}. In such a case, the dynamics of the homogeneous vector condensate $\overline {A_i}$ resembles that of the minimal scalar field and $\overline {A_i}$ can remain constant during inflation.\footnote{
	For $\alpha < -4$, $\overline A_i$ increases during inflation. At some point the backreaction of the vector boson to the inflaton becomes important, and it may lead to a scenario of so-called anisotropic inflation~\cite{Watanabe:2009ct,Soda:2012zm,Maleknejad:2012fw}. We do not pursue this possibility further in this paper.
}

Here are several comments. For $\alpha=-4$, the kinetic function $f^2 \propto a^{-4}$ is an exponentially decreasing function during inflation. It means that $f$ is exponentially large as time goes back. One might worry about the backreaction to the scalar field dynamics due to the coupling $f^2 \mathcal F \mathcal F$, but actually it is often safely neglected. Let us suppose that $\phi$ is an inflaton with its scalar potential $V(\phi)$. As we have shown above, $f^2 \mathcal F \mathcal F \sim (f \dot A_i /a)^2$ and $\overline{A_i}=fA_i /a \sim {\rm const}$ and we have $f^2 \mathcal F \mathcal F \sim H^2\overline{A_i}^2$. In order for this term not to affect the inflaton dynamics, $|(\partial_\phi f^2) \mathcal F \mathcal F| \lesssim |\partial_\phi V(\phi)|$ must be satisfied.
This condition is rewritten as 
\begin{align}
	\left(\frac{\overline{A^{(\rm in)}}}{M_{\rm P}}\right)^2 \lesssim \frac{M_{\rm P}^2}{2}\left( \frac{\partial_\phi V(\phi)}{V(\phi)} \right)^2 = \frac{r}{16}
	\sim 10^{-2} \left( \frac{H_{\rm inf}}{10^{14}\,{\rm GeV}} \right)^2,   \label{constraint_a-4}
\end{align}
where $\overline{A^{\rm (in)}}$ is a constant value of $\overline{A_i}$ during inflation, $M_{\rm P}$ denotes the reduced Planck scale, $r$ is the tensor-to-scalar ratio and $H_{\rm inf}$ is the Hubble scale during inflation~\cite{Liddle:2000cg}.\footnote{
	A similar kinetic coupling of the electromagnetic field is often considered in the context of inflationary magnetogenesis~\cite{Ratra:1991bn,Giovannini:2001nh,Bamba:2003av,Kandus:2010nw}, but it is known that models producing an observed amount of magnetic field suffer from the backreaction problem~\cite{Demozzi:2009fu,Kanno:2009ei,Fujita:2012rb}. In the magnetogenesis context it should be noticed that the magnetic field is not generated from the homogeneous $\vec A$ since $\vec B = \vec\nabla\times\vec A$ and hence the power spectrum of the magnetic field is strongly blue. In order to obtain the magnetic field in the Mpc scale, one needs huge amount of total magnetic energy or strongly red spectrum, which yield too large backreaction to the scalar field.
}
Since $H_{\rm inf}$ is bounded as $H_{\rm inf} \lesssim 10^{14}$\,GeV from the non-observation of B-mode polarization~\cite{Akrami:2018odb},
it gives an upper bound on the initial vector amplitude $\overline{A_i}$. 
%As we will see later in Sec.~\ref{sec:fluc}, the constraint from DM isocurvature perturbation gives a similar upper bound.

For $\alpha=2$, the kinetic function $f^2 \propto a^{2}$ is an exponentially increasing function during inflation. Thus $f$ is exponentially small as time goes back, which means that the theory is in a strongly coupled regime since $f$ is roughly an inverse of the gauge coupling. In a pure U(1) gauge theory without any U(1) charged field, however, there is no gauge coupling in the action and it may not cause any problem.
On the other hand, since we need $m/f \ll H$ during inflation, we require at least 
\begin{align}
	m \ll e^{-50} H_{\rm inf} \sim 10^{-22} H_{\rm inf} = 10\,{\rm eV} \left( \frac{H_{\rm inf}}{10^{14}\,{\rm GeV}} \right).
	\label{constraint_a2}
\end{align}
Thus in this case the vector boson must be very light.\footnote{
	For this reason, the case of $\alpha=2$ with constant mass $m$ was not taken seriously in the vector curvaton context since such a light curvaton is unlikely to decay before Big-Bang nucleosynthesis~\cite{Dimopoulos:2007zb}. In our case, the vector field needs not decay (actually it must not decay for it to be DM) and hence we do not discard this possibility.
}

We give one concrete example to realize a required time dependence $f^2(\phi) \propto a^\alpha$~\cite{Namba:2012gg}. Let us suppose that $\phi$ is an inflaton with power law potential $V(\phi) \sim \lambda \phi^n$. The standard slow-roll inflaton dynamics gives $\phi^2 \sim 2n N_e M_{\rm P}^2$ with $N_e$ being the e-folding number measured from the end of inflation. Thus the choice of the kinetic function
\begin{align}
	f^2(\phi) = e^{c\phi^2/M_{\rm P}^2} \simeq \left( \frac{a_{\rm end}}{a(N_e)} \right)^{2cn},   \label{f2_eg}
\end{align}
where $a_{\rm end}$ denotes the scale factor at the end of inflation and $2cn = -\alpha$. After inflation it smoothly connects to $f \to 1$.
Although a monomial inflaton potential is disfavored by the observation of the cosmic microwave background~\cite{Akrami:2018odb}, a slight modification makes the prediction of density perturbation consistent with observations~\cite{Destri:2007pv,Nakayama:2013jka,Nakayama:2013txa}. Note also that $\alpha$ needs not be exactly equal to $-4$ or $2$ for our scenario to work. What we actually need is to make the effective mass term in the equation of motion (\ref{eom_f_zero}) smaller than $\sim H^2$ during inflation.
Although we focus on the case of $\alpha=-4$ or $2$ hereafter, we can have a similar vector dynamics for more broad value of $\alpha$.

Now let us estimate the abundance of the vector coherent oscillation. As shown above, $\overline{A_i}$ remains constant during inflation for $\alpha=-4$ or $2$ and we take the initial value just as a free parameter and denote by $\overline{A^{\rm (in)}}$.\footnote{
	See Ref.~\cite{Sanchez:2013zst} for a scenario that the value of $\overline{A_i}$ is determined by the balance between the quantum fluctuation and the classical dynamics.
}
We also assume that $f=1$ after inflation and introduce an equation of state parameter $w$ until the end of the reheating, such that 
\begin{align}
	\dot H = -\frac{3}{2}(1+w) H^2,~~~~~~\frac{R}{6} = 2H^2+\dot H = \frac{1-3w}{2}H^2.  \label{R/6}
\end{align}
For a monomial power law inflaton potential $V\propto \phi^n$, it is given by $w=(n-2)/(n+2)$. The equation of motion of the zero mode after inflation then becomes
\begin{align}
	\ddot {\overline {A_i}} + 3H\dot {\overline {A_i}}  + \left(m^2 +\frac{1-3w}{2}H^2 \right) \overline{A_i} = 0.
\end{align}
Neglecting $m^2$ term, the solution looks like
\begin{align}
	\overline {A_i} \propto C_1 a^{-1} + C_2 a^{(3w-1)/2}.     \label{sol_after_inf}
\end{align}
with some numerical constants $C_1$ and $C_2$. 
%For $w=1/3$, $\overline{A_i}$ is constant until $m$ becomes comparable to $H$, after that the vector coherent oscillation begins with an amplitude $\overline{A_i^{\rm (in)}}$.
%For $w=0$, it decreases as $\overline{A_i} \propto a^{-1/2}$ until the end of the reheating, after which $w=0$ and $\overline{A_i}\sim$ const. Finally it begins a coherent oscillation at $H \sim m$.
Starting from the initial value $\overline{A_i^{\rm (in)}}$ during inflation, which is just taken as a free parameter, $\overline{A_i}$ evolves according to $C_1$ term solution or $C_2$ term solution of (\ref{sol_after_inf}). Finally it begins a coherent oscillation at $H \sim m$.\footnote{
	We assume that the reheating is completed before reaching $H=m$ since we are mainly interested in the very light DM scenario. It is justified for $m \lesssim T_{\rm R}^2 / M_{\rm P}$ with $T_{\rm R}$ being the reheating temperature.
}
We denote by $\epsilon$ the suppression factor of homogeneous vector field during this period. The vector boson begins to oscillate at $H\sim m$ with amplitude $\epsilon \overline{A_i^{\rm (in)}}$, after which it behaves as non-relativistic matter.
For example, if the $C_1$ term solution applies, then $\overline{A_i} \propto a^{-1}$ likely holds even after the reheating, hence we have
\begin{align}
	\epsilon \sim \left(\frac{H_{\rm R}}{ H_{\rm inf}}\right)^{1/2}  \left(\frac{m}{ H_{\rm R}}\right)^{\frac{2}{3(1+w)}},   \label{eps_C1}
\end{align}
where $H_{\rm R}$ denotes the Hubble scale at the end of reheating. If, on the other hand, the $C_2$ term solution of (\ref{sol_after_inf}) applies, we would have
\begin{align}
	\epsilon \sim \left(\frac{H_{\rm R}}{ H_{\rm inf}}\right)^{\frac{3w-1}{3(1+w)}},   \label{eps_C2}
\end{align}
assuming $\overline{A_i} \sim {\rm const}$ after the completion of reheating. In particular, we can have $\epsilon=1$ for $w=1/3$. 
Notice that the vector boson energy density is dominated by $\rho_A^{(K)} \sim H^2 \overline {A_i}^2/2$ for $\alpha=-4$ and the energy conservation (\ref{rhoK}) implies $\rho_A^{(K)} \propto a^{-4}$ until $H \sim m$, which is consistent with the $C_2$ term solution.
On the other hand, the $C_1$ term solution gives $\rho_A^{(K)} \sim 0$ which may be consistent with $\alpha=2$, which actually gives $\rho_A^{(K)} \sim 0$ during inflation. 
The resultant vector coherent oscillation energy density divided by the entropy density is calculated as
\begin{align}
	\frac{\rho_A}{s} = \frac{1}{8} \left(\frac{90}{\pi^2 g_*}\right)^{1/4} \frac{m^{1/2} \epsilon^2 {\overline{A^{\rm (in)}}^2} }{M_{\rm P}^{3/2}}
	\simeq 1\,{\rm GeV} \left(\frac{m}{10^{-9}\,{\rm eV}}\right)^{1/2}\left(\frac{\epsilon \overline{A^{\rm (in)}} }{M_{\rm P}}\right)^2,
\end{align}
where $g_*$ denotes the relativistic degrees of freedom at the reheating. Depending on the choice of vector boson mass and the initial condition, we can have a right amount of vector coherent oscillation DM: $\rho_{\rm DM}/s \sim 4\times 10^{-10}$\,GeV. In particular, a consistent scenario for an ultra-light vector DM such as $m\lesssim 10^{-20}\,$eV appears.

%%%%%%%%%%%%%%%%
\begin{figure}
\begin{center}
\begin{tabular}{cc}
\includegraphics[scale=1.3]{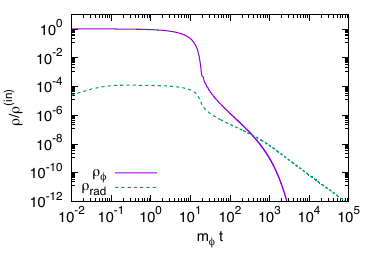}
\includegraphics[scale=1.3]{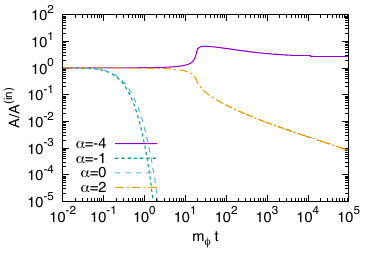}
\end{tabular}
\end{center}
\caption{
	(Left) Background evolution of inflaton energy density $\rho_\phi$ and radiation energy density $\rho_{\rm rad}$ normalized by the initial inflaton energy density $\rho^{\rm (in)}$. The inflaton decay rate is taken to be $\Gamma_\phi = 3\times 10^{-3} m_\phi$. The presence of radiation during inflation is an artifact of constant $\Gamma_\phi$ during and after inflation.
	(Right) Time evolution of physical homogeneous vector field $\overline {A_i}$, normalized by its initial value $\overline {A^{(\rm in)}}$, for different values of $\alpha$.
}
\label{fig:n2}
\end{figure}
%%%%%%%%%%%%%%%%

We performed numerical calculation to check the behavior of a vector condensate during and after inflation before the vector boson starts to oscillate.
We solved equation of motion of $\overline A_i$ (\ref{eom_f_zero}) with a concrete form of the kinetic function (\ref{f2_eg}) with $c=-\alpha/(2n)$, along with the equation of motion of the inflaton including its decay into the radiation with a decay width $\Gamma_\phi$:
\begin{align}
	&\ddot \phi + (3H+\Gamma_\phi)\dot \phi + \partial_\phi V(\phi) = 0,\\
	&\dot\rho_{\rm rad} + 4H \rho_{\rm rad} = \Gamma_\phi \rho_\phi,\\
	& 3M_{\rm P}^2 H^2 = \rho_\phi + \rho_{\rm rad},~~~~~~\rho_\phi \equiv \frac{1}{2}\dot\phi^2 + V(\phi).
\end{align}
For the inflaton potential we have taken $V(\phi)= m_\phi^2\phi^2/2$. 
Numerical results are shown in Fig.~\ref{fig:n2}. The left panel shows the background evolution of inflaton energy density $\rho_\phi$ and radiation energy density $\rho_{\rm rad}$ normalized by the initial inflaton energy density $\rho^{\rm (in)}$. We have taken $\phi = 15 M_{\rm P}$ at the beginning of numerical calculation. The inflaton decay rate is taken to be $\Gamma_\phi = 3\times 10^{-3} m_\phi$. The presence of radiation during inflation is an artifact of constant $\Gamma_\phi$ during and after inflation. The right panel shows time evolution of physical homogeneous vector field $\overline {A_i}$, normalized by its initial value $\overline {A^{(\rm in)}}$, for different values of $\alpha$. It is seen that  $\overline {A_i}$ exponentially decays for $\alpha \neq -4$ or $2$.
For $\alpha=-4$, the field value is constant during inflation as expected and there is a small increase around the end of inflation. This is because the effective mass square in (\ref{eom_f_zero}) temporally becomes negative during the first half inflaton oscillation. After that, the oscillation averaged mass square becomes positive as given in (\ref{R/6}). It is checked that during the inflaton matter domination $\overline {A_i} \propto a^{-1/2}$ and during the radiation domination $\overline {A_i} \sim {\rm const}$, corresponding to the $C_2$ term solution of (\ref{sol_after_inf}). Thus $\epsilon$ is evaluated by (\ref{eps_C2}) in this case.
For $\alpha=2$, the field value is constant during inflation as expected but after inflation it connects to the $C_1$ term solution of (\ref{sol_after_inf}). Thus $\overline {A_i}$ decreases as $\overline {A_i} \propto a^{-1}$ in the radiation dominated era and $\epsilon$ is evaluated by (\ref{eps_C1}).
%We are not confident whether this feature is generic or just applied to this particular example. Anyway it implies that one must be careful about which solution of (\ref{sol_after_inf}) is realized depending on the model in order to evaluate the abundance of vector coherent oscillation.

%%%%%%%%%%%%%%%%%%%%%%%%%%%
\subsection{Isocurvature fluctuation}  \label{sec:fluc}
%%%%%%%%%%%%%%%%%%%%%%%%%%%

In this class of scenario, one must be careful about the constraint from the isocurvature fluctuation of DM.
The action of the transverse mode (\ref{ST_f}) is written as
\begin{align}
	S_T = \int \frac{d\tau d^3k}{(2\pi)^3} \frac{1}{2}
	\left[ |\vec{A}^{c \prime}_T(k)|^2 - \left(k^2+\frac{a^2m^2}{f^2}-2\mathcal H^2 \right) |\vec{A}^c_T(k)|^2 \right],
\end{align}
after substituting $f^2\propto a^{\alpha}$ with $\alpha=-4$ or $2$ during inflation. Assuming $m/f\ll H$, this action is the same as the light minimal scalar field and hence the generation and evolution of the fluctuations are also the same.\footnote{
	To be precise, it is the same as the minimal scalar $\chi$ after the rescaling $\chi\to a \chi$. Recalling that it is $\vec{A}^{c}_T/a$ that may be regarded as a ``physical'' field, we can understand the correspondence between  $\chi$ and $\vec{A}^{c}_T/a$.
}
As is well known in the context of a scalar curvaton, the power spectrum has a nearly scale invariant spectrum.
The power spectrum of the transverse fluctuations is given by
\begin{align}
	\left< \vec{A}^{c}_T(k) \vec{A}^{c*}_T(k') \right>\equiv \frac{4\pi^2 a^2}{k^3}\mathcal P_T(k) (2\pi)^3 \delta(\vec k-\vec k'),~~~~~
	\mathcal P_{T}(k)\simeq \left( \frac{H_{\rm inf}}{2\pi} \right)^2.
\end{align}
Noting that $\left< \overline{A_T}^2(\vec x)\right> \simeq \int d\ln k\,\mathcal P_T(k)$, $\sqrt{\mathcal P_T(k)}$ denotes the typical amplitude of the transverse fluctuation with wavenumber $k$.

Next let us consider the longitudinal mode. The longitudinal mode action (\ref{SL_f}) is written as
\begin{align}
	S_L = \int \frac{d\tau d^3k}{(2\pi)^3} \frac{1}{2}
	\left[ |{A}^{c \prime}_L(k)|^2 - \left(k^2+\frac{a^2m^2}{f^2}-2\mathcal H^2 + \frac{27 a^2m^2 f^2k^2}{(f^2k^2+a^2m^2)^2}\mathcal H^2 \right) |{A}^c_L(k)|^2 \right],
	\label{SL_alpha-4}
\end{align}
after substituting $f^2\propto a^{\alpha}$ with $\alpha=-4$ during inflation, while
\begin{align}
	S_L = \int \frac{d\tau d^3k}{(2\pi)^3} \frac{1}{2}
	\left[ |{A}^{c \prime}_L(k)|^2 - \left(k^2+\frac{a^2m^2}{f^2}-2\mathcal H^2\right) |{A}^c_L(k)|^2 \right], 
\end{align}
after substituting $f^2\propto a^{\alpha}$ with $\alpha=2$ during inflation. For $\alpha=2$, the longitudinal mode action is the same as the transverse mode and hence we obtain the same longitudinal power spectrum as the transverse one.
For $\alpha=-4$, there is an additional term proportional to $\mathcal H^2$. Although this additional term is much smaller than the $-2\mathcal H^2$ term except for a small time interval around which $k/a\sim m/f$, one should be careful about the evolution across this intermediate epoch.
As shown in App.~\ref{app:evo}, the growing mode at $k/a\gg m/f$ connects to the decaying mode at $k/a \ll m/f$. 
As a result, the power spectrum of the longitudinal fluctuation is given as
\begin{align}
	\left< {A}^{c}_L(k) {A}^{c*}_L(k') \right>\equiv \frac{2\pi^2 a^2}{k^3}\mathcal P_L(k) (2\pi)^3 \delta(\vec k-\vec k'),
\end{align}
where
\begin{align}
	\mathcal P_{L}(k)\simeq 
	\begin{cases}
	\displaystyle\left( \frac{H_{\rm inf}}{2\pi} \right)^2 \left(\frac{k}{m a_{\rm end}}\right)^{2}  &{\rm for}~~~\alpha=-4,\\
	\displaystyle\left( \frac{H_{\rm inf}}{2\pi} \right)^2  &{\rm for}~~~\alpha=2.
	\end{cases}
\end{align}
The power spectrum is strongly blue tilted for $\alpha=-4$ and hence the longitudinal fluctuation is negligibly small compared with the transverse mode on the present cosmological scale. For $\alpha=2$, the transverse and longitudinal modes have comparable power at all the observable scale.
In both cases, the evolution of the long wavelength mode after inflation is the same as the zero mode, and hence we have the density fluctuation of the vector DM as 
\begin{align}
	\frac{\delta \rho_{A}}{\rho_A} \sim \frac{H_{\rm inf}}{\pi\overline {A^{(\rm in)}}}.
\end{align}

Since the vector fluctuation is independent of the inflaton fluctuation, their density fluctuation result in the DM isocurvature fluctuation, which is severely constrained from the cosmological observation. 
Let $\zeta(\vec x)$ be the curvature perturbation on the uniform-density time slice $\delta \rho_{\rm total}(\vec x)=0$~\cite{Lyth:2004gb}. We also define $\zeta_{\rm DM}(\vec x)$ by the curvature perturbation on the slice where $\delta\rho_{A}(\vec x)=0$ and 
\begin{align}
	\zeta_{\rm DM}(\vec x) - \zeta(\vec x) = \Delta N(\vec x) =\frac{\delta \rho_A}{3 \rho_{A}},
\end{align}
where $\Delta N$ is the e-folding number from the uniform-density slice to the $\delta\rho_{A}(\vec x)=0$ slice and $\delta \rho_{A}$  in the most right hand side is evaluated on the uniform-density slice. The non-linear DM isocurvature perturbation is defined by~\cite{Wands:2000dp,Kawasaki:2008sn,Langlois:2008vk}
\begin{align}
	S_{\rm DM} \equiv 3(\zeta_{\rm DM}-\zeta) =\frac{\delta \rho_A}{\rho_{A}} \sim \frac{H_{\rm inf}}{\pi\overline {A^{(\rm in)}}}.
	\label{S_DM}
\end{align}
In our scenario it is nearly scale invariant and it is constrained as $S_{\rm DM} \lesssim 9\times 10^{-6}$ for uncorrelated DM isocurvature perturbation~\cite{Akrami:2018odb}. 
%Remarkably this gives a similar constraint to (\ref{constraint_a-4}) for $\alpha=-4$.
Note that the Planck constraint~\cite{Akrami:2018odb} is derived by assuming that the DM isocurvature perturbation is statistically isotropic, but in our case it is anisotropic as we see below. Taking account of the statistical anisotropy, the constraint may change but it does not affect our order-of-magnitude estimate.
Remarkably this constraint has the same parameter dependence as (\ref{constraint_a-4}) for $\alpha=-4$ but with inverse inequality, and they cannot be consistent with each other, although the violation of the backreaction constraint (\ref{constraint_a-4}) does not immediately mean that such parameter regions are excluded.\footnote{
		{\bf [Note added in the latest arXiv version (Mar.2023)]} In the old version, we misinterpreted the inequality of (\ref{constraint_a-4}) and concluded that it gives ``similar'' constraint to the isocurvature bound~(\ref{S_DM}). We thank Kunio Kaneta for pointing out it. The scenario with significant backreaction has been analyzed in Ref.~\cite{Nakayama:2020rka} with a conclusion that the statistical anisotropy of the curvature perturbation becomes too large to be consistent with observations.
}

Now we point out that the DM isocurvature fluctuation in our scenario may be statistically anisotropic. It is expanded in terms of the vector field fluctuation as
\begin{align}
	S_{\rm DM} = \frac{\delta \rho_A}{\rho_A} \simeq \frac{1}{\rho_A} \frac{\partial\,\delta\rho_A}{\partial A_i} \delta A_i \equiv n^i \delta A_i.
\end{align}
By using the following decomposition~\cite{Dimopoulos:2008yv}
\begin{align}
	\left< {A}^{c}_i(\vec k) {A}^{c*}_j(\vec k') \right>\equiv \frac{2\pi^2 a^2}{k^3} (2\pi)^3 \delta(\vec k-\vec k') \left[ (\delta_{ij}-\hat k_i \hat k_j) \mathcal P_T(k) + \hat k_i \hat k_j \mathcal P_L(k) \right],
\end{align}
we find the dimensionless power spectrum of the DM isocurvature fluctuation as
\begin{align}
	\mathcal P_{S_{\rm DM}}(\vec k) &= (n^in_i)\, \mathcal P_T(k) + (n^i \hat k_i)^2(\mathcal P_L(k)-\mathcal P_T(k)) \\
	&\equiv \mathcal P^{\rm (iso)}_{S_{\rm DM}}(k) \left[ 1+  (\hat n^i \hat k_i)^2\,g_{S}(k) \right].
\end{align}
The first term is the isotropic component of the DM isocurvature perturbation and $g_S(k)= (\mathcal P_L(k)-\mathcal P_T(k))/\mathcal P_T(k)$ measures the size of statistical anisotropy of the DM isocurvature perturbation.
As shown above, the longitudinal power is negligibly small at large scale for $\alpha=-4$, which means that DM isocurvature fluctuation is dominantly anisotropic ($g_S(k)\simeq -1$). 
Moreover, one can have a blue-tilted isocurvature power spectrum by assuming a time-dependent mass term $m\propto a^\beta(t)$ with some constant $\beta$ during inflation as extensively studied in the context of vector curvaton~\cite{Dimopoulos:2007zb,Dimopoulos:2009am,Dimopoulos:2009vu} and also shown in App.~\ref{app:mass}, which may enhance the chance to detect isocurvature fluctuation at relatively small scale.\footnote{
	In the vector curvaton scenario of Refs.~\cite{Dimopoulos:2007zb,Dimopoulos:2009am,Dimopoulos:2009vu}, the curvature perturbation is assumed to be sourced by the vector field, and hence three conditions are required: the (nearly) scale-invariant spectrum, the cancellation of the statistical anisotropy (which implies $\mathcal P_T\simeq \mathcal P_L$ at large scale) and decay of the curvaton before Big-Bang nucleosynthesis. For $\alpha=-4$, the second condition is satisfied for $\beta=1$. For $\alpha=2$, although $\beta=0$ satisfies the first and second condition, it is difficult to achieve the last condition. All the conditions are satisfied for $\beta=1$. See App.~\ref{app:mass}.
}
This statistically anisotropic DM isocurvature fluctuation can be a unique signal to distinguish vector coherent oscillation DM from other DM candidates.
%Although we have obtained a similar size of the transverse and longitudinal fluctuation of power, they need not exactly coincide with each other. As extensively studied in the context of vector curvaton~\cite{Dimopoulos:2007zb,Dimopoulos:2009am,Dimopoulos:2009vu}, one can assume time-dependent mass term $m\propto a^\beta(t)$ with some constant $\beta$ during inflation, as far as it does not spoil the condition $m(t)/f(t) \ll H_{\rm inf}$. In such a case the longitudinal mode can have a very different spectrum, which would enhance the statistical anisotropy in the DM isocurvature fluctuation.

\begin{figure}[t]
\begin{center}
\includegraphics[scale=1.8]{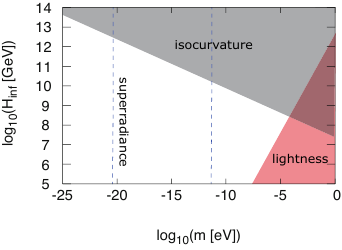}
\end{center}
\caption{The allowed parameter region for the vector coherent DM model on the plane of $(m,H_{\rm inf})$. 
At each value of $m$, $\overline{A^{(\rm in)}}$ is chosen such that the vector coherent oscillation becomes dominant component of DM for $\epsilon=1$. The upper gray shaded region is excluded by the DM isocurvature constraint. 
Although not explicitly shown in this figure, the backreaction constraint (\ref{constraint_a-4}) is complementary to the isocurvature constraint.
The lower right red shaded region is excluded by the constraint (\ref{constraint_a2}), which applies to the case of $\alpha=2$. The region sandwiched by two vertical dashed blue lines may be explored through the black hole superradiance.
}
\label{fig:cont}
\end{figure}

Fig.~\ref{fig:cont} summarizes constraints on our vector coherent DM scenario.\footnote{	{\bf [Note added in the latest arXiv version (Mar.2023)]} One should note that it is very difficult to have $\epsilon=1$ for $\alpha=2$. With a realistic choice of $\epsilon$, the $\alpha=2$ case may not able to explain the observed DM abundance. If we also introduce a coupling like $\sim f^2(\phi) m^2\mathcal A_M \mathcal A^M$, it is rather easy to obtain a correct relic abundance, while the isocurvature perturbation of the longitudinal mode is greatly enhanced~\cite{Nakayama:2020rka}.  }
At each value of $m$, $\overline{A^{(\rm in)}}$ is chosen such that the vector coherent oscillation becomes dominant component of DM for $\epsilon=1$. The upper gray shaded region is excluded from the DM isocurvature constraint. 
Although not explicitly shown in this figure, the backreaction constraint (\ref{constraint_a-4}) is complementary to the isocurvature constraint.
The lower right red shaded region is excluded by the constraint (\ref{constraint_a2}), which applies only to the case of $\alpha=2$, although it is not clear whether we can have $\epsilon=1$ for $\alpha=2$ or not. The region sandwiched by two vertical dashed blue lines may be explored through the black hole superradiance~\cite{Arvanitaki:2009fg,Arvanitaki:2010sy,Pani:2012vp,Brito:2015oca}.

%%%%%%%%%%%%%%%%
%%%%%%%%%%%%%%%%

%%%%%%%%%%%%%%%%%%%%%%%%%%%%%%%%%%%%%%%%%%%%%%%%%%
\section{Discussion} \label{sec:dis}
%%%%%%%%%%%%%%%%%%%%%%%%%%%%%%%%%%%%%%%%%%%%%%%%%%

In this paper we constructed a model of hidden massive vector DM as a homogeneous coherent oscillation.
We introduced a scalar dependent gauge kinetic function $f^2(\phi)$ in order to break the conformal invariance and to persist a vector field condensate during inflation. Our model does not suffer from neither ghost nor gradient instabilities, while some specific time dependence of the kinetic function is mandatory for this purpose. As a result we obtained a consistent scenario for an ultra-light vector DM as coherent oscillation.\footnote{
		{\bf [Note added in the latest arXiv version (Mar.2023)]} Actually the case of $\alpha=-4$ requires careful treatments of the from the backreation and the $\alpha=2$ case is also extremely difficult to obtain the correct relic DM abundance. See Ref.~\cite{Nakayama:2020rka} for more detailed discussion.}
The evolution of the long wave fluctuations are also calculated and it is pointed out that the DM isocurvature perturbation has a distinct property that it can be highly statistically anisotropic, which can be a smoking-gun signal of a vector coherent oscillation DM scenario.

Here let us discuss the detection possibility.
As we have just mentioned, the discovery of the statistically anisotropic DM isocurvature perturbation could be a unique signal of vector coherent DM. 
The constraint on the uncorrelated DM isocurvature perturbation $\mathcal P_{S_{\rm DM}}$ will be improved by a factor five at the future CMB-S4 experiment~\cite{Abazajian:2016yjj} compared with the current constraint.
Although the detailed estimation of the sensitivity on the statistically anisotropic DM isocurvature perturbation at the current and future experiments is beyond the scope of this paper, a similar improvement is naively expected. Furthermore, in our scenario the isocurvature perturbation can naturally have a strongly blue spectrum, which implies that it tends to affect the small scale CMB anisotropy and hence it is more sensitive to the precise cosmological observations at the small scale compared with usual scale-independent isocurvature perturbation.
Light vector boson with $10^{-20}\,{\rm eV}\lesssim m \lesssim 10^{-11}$\,eV may also induce a superradiance instability of the astrophysical black holes~\cite{Arvanitaki:2009fg,Arvanitaki:2010sy,Pani:2012vp,Brito:2015oca}. Observations of spinning black holes can constrain such mass regions.
One can also introduce a kinetic mixing of the hidden vector boson to the hypercharge photon,
\begin{align}
	\mathcal L = -\frac{\kappa}{2} F^{(Y)}_{MN} F^{MN},
\end{align}
where $\kappa$ is a constant and $F_{MN}^{(Y)}$ denotes the field strength of the hypercharge photon. This kinetic mixing provides us with lots of phenomenology and opens up a possibility to detect it through terrestrial experiments~\cite{Jaeckel:2010ni,Arias:2012az,Horns:2012jf,Parker:2013fxa,Chaudhuri:2014dla,Hochberg:2016ajh,Hochberg:2016sqx,Arvanitaki:2017nhi,Baryakhtar:2018doz}.
Note that the effective kinetic mixing parameter is rather given by $\kappa_{\rm eff}\equiv \kappa / f(\phi)$ during inflation. For $\alpha=-4$, $\kappa_{\rm eff}$ becomes smaller at earlier epoch. For $\alpha=2$, it becomes larger and larger for earlier epoch hence it may enter a strong couple regime. To avoid this, one can also assume time dependent $\kappa$ by regarding it as a function of some scalar field as $\kappa(\phi)$.

Finally we mention other possibilities to have a homogeneous vector condensate. Actually in the context of inflaionary model building it is known that there are several situations where the vector condensate plays a crucial role. 
In a so-called gauge-flation scenario~\cite{Maleknejad:2011jw,Maleknejad:2011sq,Maleknejad:2012fw}, hidden SU(2) gauge fields obtain homogeneous field values causing inflation while the three gauge bosons are aligned with the $(x,y,z)$ direction respectively so that the isotropic expansion of the universe is retained. A crucial difference between U(1) and non-Abelian gauge boson is that the latter necessarily has self interactions without violating gauge invariance, which may serve as a potential of the gauge field, and also one can introduce a term like $\mathcal L \sim (F_{MN}^a \widetilde F^{MN a})^2$ to modify the kinetic term of the background homogeneous gauge boson.
Such an idea may be used to construct a consistent model of vector coherent oscillation DM, although the gauge coupling constant should be extremely small to satisfy the bound on the DM self interaction for an ultra-light DM.
Another scenario is a so-called anisotropic inflation~\cite{Watanabe:2009ct,Maleknejad:2012fw,Soda:2012zm}. In this scenario gauge kinetic function of $f^2(\phi) F_{MN}F^{MN}$ is introduced and the homogeneous vector field is supported by the time dependence of $f$, similar to our DM scenario, while it backreacts to the inflaton dynamics so that it has another slow-roll anisotropic inflation phase.
As briefly mentioned in Sec.~\ref{sec:zero}, this may corresponds to the case $\alpha<-4$ and it might be interesting to pursue this possibility as a mechanism to create vector coherent oscillation DM.

%%%%%%%%%%%%%%%%%%%%%%%%%%%%%%%%%%%%%%%%%%%%
\section*{Acknowledgments}
%%%%%%%%%%%%%%%%%%%%%%%%%%%%%%%%%%%%%%%%%%%%

This work was supported by the Grant-in-Aid for Scientific Research C (No.18K03609 [KN]) and Innovative Areas (No.15H05888 [KN], No.17H06359 [KN]).

%%%%%%%%%%%%%%%%%%%%%%%%%%%%%%%%%%%%%%%%%%%%%%%%%%
\appendix
%%%%%%%%%%%%%%%%%%%%%%%%%%%%%%%%%%%%%%%%%%%%%%%%%%
\section{Evolution of mode function} \label{app:evo}
%%%%%%%%%%%%%%%%%%%%%%%%%%%%%%%%%%%%%%%%%%%%%%%%%%

In this Appendix we give a solution to the equation of motion derived from the longitudinal mode action for $\alpha=-4$ (\ref{SL_alpha-4}). Especially we are interested in the evolution across the epoch of $k/a \sim m/f$.
The mode function $\widetilde A^c_L(\vec k,\tau) $ is defined by
\begin{align}
	A_L^{c}(\vec k,\tau) = \widetilde A^c_L(\vec k,\tau) a_{\vec k} + \widetilde A^{c*}_L(\vec k,\tau) a^\dagger_{-\vec k}, 
\end{align}
where the creation and annihilation operators satisfy
\begin{align}
	\left[ a_{\vec k},a^\dagger_{\vec k'}\right] = (2\pi)^3 \delta(\vec k-\vec k'),~~~~~~
	\left[ a_{\vec k},a_{\vec k'}\right] =\left[ a^\dagger_{\vec k},a^\dagger_{\vec k'}\right] = 0.
\end{align}
The vacuum state is defined such that $a_{\vec k} \left|0\right> = 0$.
Using this mode function, the power spectrum is given by
\begin{align}
	\left< {A}^{c}_L(\vec k) {A}^{c*}_L(\vec k') \right> = \left|\widetilde A^c_L(k)\right|^2 (2\pi)^3 \delta(\vec k-\vec k').
\end{align}

First, for $k/a \gg m/f$, the equation is the same as the transverse mode, which is nothing but a minimally coupled scalar as already mentioned in the main text. Taking the Bunch-Davis boundary condition in the short wavelength limit, we have
\begin{align}
	\widetilde A_L^{c}(k,\tau) \simeq \frac{1}{\sqrt{2k}} \sqrt{\frac{-\pi k\tau}{2}} H_{3/2}^{(1)}(-k\tau)
	\simeq \begin{cases}
		\displaystyle\frac{1}{\sqrt{2k}} e^{-ik\tau} & {\rm for}~~~k/a\gg H_{\rm inf},\\
		\displaystyle\frac{a H_{\rm inf}}{\sqrt{2} k^{3/2}}  & {\rm for}~~~m/f\ll k/a\ll H_{\rm inf},
	\end{cases}
\end{align}
where $H_{3/2}^{(1)}(z)$ denotes the Hankel function of the first kind. Thus in the superhorizon regime $A_L^c$ starts with the growing solution $\propto a$. Recall that the physical field is $A_L^c/a$, which remains constant in this regime and hence the physical energy density also remains constant.

Next let us consider the evolution across the epoch of $k/a \sim m/f$.
By using $\tau= -1/(aH_{\rm inf})$ during inflation, the equation of motion for the superhorizon mode $k/a \ll H_{\rm inf}$ and for $m/f \ll H_{\rm inf}$ is written in the form of
\begin{align}
	\widetilde A_L^{c \prime\prime}(k,\tau) - \left( \frac{2}{\tau^2} - \frac{27\tau^4\tau_*^6}{(\tau^6+\tau_*^6)^2} \right)\widetilde A_L^c(k,\tau) = 0.
\end{align}
Here $\tau < \tau_*$ and $\tau > \tau_*$ correspond to $k/a < m/f$ and $k/a > m/f$, respectively. One easily finds that in both the regime $\tau \ll \tau_*$ and $\tau \gg \tau_*$, a solution looks like $\widetilde A_L^c \sim C_1\tau^2 + C_2 \tau^{-1}$ with some constants $C_1$ and $C_2$.
Thus one may naively expect that the growing solution at $\tau \ll \tau_*$ smoothly connects to the growing solution at $\tau \gg \tau_*$. This is not true, however. The exact solution is given by
\begin{align}
	\widetilde A_L^{c}(k,\tau) = C_1 \frac{\tau^2}{(\tau^6+\tau_*^6)^{1/2}} + C_2\frac{\tau^6-\tau_*^6}{\tau(\tau^6+\tau_*^6)^{1/2}},
\end{align}
with some constants $C_1$ and $C_2$. One sees that the $C_1$ term connects the initial growing solution to the final decaying solution and $C_2$ term connects the initial decaying solution to the final growing solution. Since we already obtained a growing solution at $m/f\ll k$, we end up with the decaying solution.

Therefore, at the end of inflation, the power spectrum for $k\lesssim a_{\rm end} m$ is given as
\begin{align}
	\left|\widetilde A^c_L(k,t_{\rm end})\right|^2 \sim \frac{1}{2k}\left( \frac{a_* H_{\rm inf}}{k} \right)^2 \left( \frac{a_*}{a_{\rm end}} \right)^4 \sim
	 \frac{a_{\rm end}^2 H_{\rm inf}^2}{2k^3}\left(\frac{k}{m a_{\rm end}}\right)^{2},
\end{align}
where $a_*$ denotes the scale factor at $k/a = m/f$ and we have taken $f=1$ at the end of inflation. Thus the power spectrum is strongly blue tilted. At the observable scale, i.e. present cosmological scale, the longitudinal fluctuation is negligibly small compared with the transverse mode.

%%%%%%%%%%%%%%%%%%%%%%%%%%%%%%%%%%%%%%%%%%%%%%%%%%
\section{Varying mass term} \label{app:mass}
%%%%%%%%%%%%%%%%%%%%%%%%%%%%%%%%%%%%%%%%%%%%%%%%%%

In this Appendix we mention the case of time-dependent vector boson mass term $m(t) \propto a^\beta(t)$ with some constant $\beta$ during inflation. For simplicity we assume $m$ approaches to $m_0$ at the end of inflation and it remains constant thereafter.
This possibility was considered in the context of vector curvaton~\cite{Dimopoulos:2007zb,Dimopoulos:2009am,Dimopoulos:2009vu} and it is worth mentioning how their results and ours are related.

The zero mode dynamics is not affected by the time dependence of the mass term as far as $m/f\ll H$ is satisfied during inflation.
Similarly, the transverse mode action (\ref{ST_f}) is not affected. It is the longitudinal fluctuation dynamics that can be significantly modified. The effective mass term of the longitudinal vector boson, $-g''/g$ in (\ref{SL_f}), is now given by  
\begin{align}
	\frac{g''}{g} = \frac{ \mathcal H^2}{4}
	\frac{4 (1+\beta)(2+\beta)f^4k^4 + \alpha(2+\alpha)a^4m^4 -2(2-7\alpha +\alpha^2+6\beta-6\alpha\beta+4\beta^2) f^2k^2a^2m^2}
	{(f^2k^2+a^2m^2)^2},
\end{align}
where we have assumed inflationary epoch during which $a''/a=2\mathcal H^2$. The limiting form in the short and long wavelength limit is given by
\begin{align}
	\frac{g''}{g} \to \begin{cases}
		(1+\beta)(2+\beta) \mathcal H^2 & {\rm for}~~~fk \gg am,\\
		\displaystyle \frac{\alpha(2+\alpha)}{4} \mathcal H^2 & {\rm for}~~~fk \ll am.
	\end{cases}
\end{align}
In our model studied in the main text, $\beta=0$ and $\alpha=-4$ or $2$ are assumed and we have $g''/g \to 2\mathcal H^2$ in the both limit.
%Recalling that $-g''/g$ gives an effective mass for the longitudinal vector boson (see Eq.~(\ref{SL_f})), the transverse and longitudinal modes behave in a similar way except for a short transition epoch around $fk\sim am$. 
In the vector curvaton model studied in Ref.~\cite{Dimopoulos:2009am}, on the other hand, it is assumed that $\beta=1$ with $\alpha=-4$ or $2$. In this case, we have
\begin{align}
	\frac{g''}{g} \to \begin{cases}
		 6\mathcal H^2 & {\rm for}~~~fk \gg am,\\
		 2\mathcal H^2 & {\rm for}~~~fk \ll am.
	\end{cases}
\end{align}
Thus the superhorizon evolution of the longitudinal mode ($m/f \ll k/a \ll H$) is different from the transverse one.
Keeping general value of $\beta$, the solution of the equation of motion in each regime during inflation is schematically given as follows:\footnote{
	The equation of motion in the superhorizon regime is approximated by the form $\widetilde A_L^{c\prime\prime} - n\mathcal H^2 \widetilde A_L^c=0$, whose solution is given by $\widetilde A_L^c \propto a^{-(1\pm\sqrt{1+4n})/2}$ during inflation. In particular, in the case of $n=(1+\beta)(2+\beta)$, the solution is $\widetilde A_L^c \propto a^{-(2+\beta)}$ and $a^{1+\beta}$.
}
\begin{align}
	\widetilde A^c_L(k,t) \sim \begin{cases}
	 \displaystyle\frac{1}{\sqrt{2k}} e^{-ik\tau}&{\rm for}~~~k/a \ll H,\\
         c_1 a^{1+\beta} + c_2 a^{-(2+\beta)}&{\rm for}~~~m/f \ll k/a \ll H,\\
	c_3 a + c_4 a^{-2} &{\rm for}~~~k/a \ll m/f.
	\end{cases}
\end{align}
The unusual growing solution in the intermediate regime at $m/f \ll k/a \ll H$ modifies the resulting spectrum of the fluctuation. 
As noted in App.~\ref{app:evo}, it is non-trivial whether the mode function finally enters the growing solution $(\propto a)$ or decaying solution ($\propto a^{-2}$).
If the growing solution applies, which is actually the case for $\alpha=2$, the superhorizon spectrum for $k/a < m/f$ at the end of inflation looks like
\begin{align}
	\left|\widetilde A^c_L(k,t_{\rm end})\right|^2 \sim \frac{a_{\rm end}^2 H_{\rm inf}^2}{2k^3}\left(\frac{H_{\rm inf}}{m_0}\right)^{2\beta}
	\left(\frac{k}{m_0 a_{\rm end}}\right)^{\frac{2\beta(\alpha-2\beta)}{2+2\beta-\alpha}},
\end{align}
where we have taken $f=1$ and $m=m_0$ at the end of inflation. Thus the power spectrum deviates from the scale invariant one in general. The scale invariance is recovered for $\beta=0$ or $\alpha-2\beta=0$, the latter of which corresponds to $(\alpha,\beta)=(2,1)$ as extensively studied in the vector curvaton context~\cite{Dimopoulos:2009am}.
If the decaying solution applies, which is actually the case for $\alpha=-4$, the superhorizon spectrum for $k/a < m/f$ at the end of inflation, looks like
\begin{align}
	\left|\widetilde A^c_L(k,t_{\rm end})\right|^2 \sim \frac{a_{\rm end}^2 H_{\rm inf}^2}{2k^3}\left(\frac{H_{\rm inf}}{m_0}\right)^{2\beta}
	\left(\frac{k}{m_0 a_{\rm end}}\right)^{\frac{12+2\beta(\alpha-2\beta)}{2+2\beta-\alpha}},
\end{align}
Again the power spectrum deviates from the scale invariant one in general. The scale invariance is recovered for $6+\beta(\alpha-2\beta)=0$, which corresponds to $(\alpha,\beta)=(-4,1)$ as also extensively studied in the vector curvaton context~\cite{Dimopoulos:2009am}.
In our scenario to identify the vector boson as DM, these kind of modified spectrum are imprinted in the DM isocurvature perturbation and it may be interesting that the blue spectrum enhances the detection possibility.

%%%%%%%%%%%%%%%%%%%%%%%%%%%%%%%%%%%%%%%%%%%%%%%%%%

%%%%%%%%%%%%%%%%%%%%%%%%%%%%%%%%%%%%%%%%%%%%%%%%%%


\begin{thebibliography}{99}
%%%%%%%%%%%%%%%%%%%%%%%%%%%%%%%%%%%%%%%%%%%%%%%%%%


%\cite{Jaeckel:2010ni}
\bibitem{Jaeckel:2010ni} 
  J.~Jaeckel and A.~Ringwald,
  %``The Low-Energy Frontier of Particle Physics,''
  Ann.\ Rev.\ Nucl.\ Part.\ Sci.\  {\bf 60}, 405 (2010)
  %doi:10.1146/annurev.nucl.012809.104433
  [arXiv:1002.0329 [hep-ph]].
  %%CITATION = doi:10.1146/annurev.nucl.012809.104433;%%

 %\cite{Arias:2012az}
\bibitem{Arias:2012az} 
  P.~Arias, D.~Cadamuro, M.~Goodsell, J.~Jaeckel, J.~Redondo and A.~Ringwald,
  %``WISPy Cold Dark Matter,''
  JCAP {\bf 1206}, 013 (2012)
  %doi:10.1088/1475-7516/2012/06/013
  [arXiv:1201.5902 [hep-ph]].
  %%CITATION = doi:10.1088/1475-7516/2012/06/013;%%
  
  %\cite{Preskill:1982cy}
\bibitem{Preskill:1982cy} 
  J.~Preskill, M.~B.~Wise and F.~Wilczek,
  %``Cosmology of the Invisible Axion,''
  Phys.\ Lett.\ B {\bf 120}, 127 (1983).
  %[Phys.\ Lett.\  {\bf 120B}, 127 (1983)].
  %doi:10.1016/0370-2693(83)90637-8
  %%CITATION = doi:10.1016/0370-2693(83)90637-8;%%
  
  %\cite{Abbott:1982af}
\bibitem{Abbott:1982af} 
  L.~F.~Abbott and P.~Sikivie,
  %``A Cosmological Bound on the Invisible Axion,''
  Phys.\ Lett.\ B {\bf 120}, 133 (1983).
  %[Phys.\ Lett.\  {\bf 120B}, 133 (1983)].
  %doi:10.1016/0370-2693(83)90638-X
  %%CITATION = doi:10.1016/0370-2693(83)90638-X;%%
  
  %\cite{Dine:1982ah}
\bibitem{Dine:1982ah} 
  M.~Dine and W.~Fischler,
  %``The Not So Harmless Axion,''
  Phys.\ Lett.\ B {\bf 120}, 137 (1983).
  %[Phys.\ Lett.\  {\bf 120B}, 137 (1983)].
  %doi:10.1016/0370-2693(83)90639-1
  %%CITATION = doi:10.1016/0370-2693(83)90639-1;%%
  
  %\cite{Kim:1986ax}
\bibitem{Kim:1986ax} 
  J.~E.~Kim,
  %``Light Pseudoscalars, Particle Physics and Cosmology,''
  Phys.\ Rept.\  {\bf 150}, 1 (1987).
  %doi:10.1016/0370-1573(87)90017-2
  %%CITATION = doi:10.1016/0370-1573(87)90017-2;%%
  
  %\cite{Kawasaki:2013ae}
\bibitem{Kawasaki:2013ae} 
  M.~Kawasaki and K.~Nakayama,
  %``Axions: Theory and Cosmological Role,''
  Ann.\ Rev.\ Nucl.\ Part.\ Sci.\  {\bf 63}, 69 (2013)
  %doi:10.1146/annurev-nucl-102212-170536
  [arXiv:1301.1123 [hep-ph]].
  %%CITATION = doi:10.1146/annurev-nucl-102212-170536;%%
  
  
  
  %\cite{Agrawal:2018vin}
\bibitem{Agrawal:2018vin} 
  P.~Agrawal, N.~Kitajima, M.~Reece, T.~Sekiguchi and F.~Takahashi,
  %``Relic Abundance of Dark Photon Dark Matter,''
  arXiv:1810.07188 [hep-ph].
  %%CITATION = ARXIV:1810.07188;%%
  
  %\cite{Co:2018lka}
\bibitem{Co:2018lka} 
  R.~T.~Co, A.~Pierce, Z.~Zhang and Y.~Zhao,
  %``Dark Photon Dark Matter Produced by Axion Oscillations,''
  Phys.\ Rev.\ D {\bf 99}, no. 7, 075002 (2019)
  %doi:10.1103/PhysRevD.99.075002
  [arXiv:1810.07196 [hep-ph]].
  %%CITATION = doi:10.1103/PhysRevD.99.075002;%%
  
  %\cite{Bastero-Gil:2018uel}
\bibitem{Bastero-Gil:2018uel} 
  M.~Bastero-Gil, J.~Santiago, L.~Ubaldi and R.~Vega-Morales,
  %``Vector dark matter production at the end of inflation,''
  JCAP {\bf 1904}, no. 04, 015 (2019)
  %doi:10.1088/1475-7516/2019/04/015
  [arXiv:1810.07208 [hep-ph]].
  %%CITATION = doi:10.1088/1475-7516/2019/04/015;%%

%\cite{Dror:2018pdh}
\bibitem{Dror:2018pdh} 
  J.~A.~Dror, K.~Harigaya and V.~Narayan,
  %``Parametric Resonance Production of Ultralight Vector Dark Matter,''
  Phys.\ Rev.\ D {\bf 99}, no. 3, 035036 (2019)
  %doi:10.1103/PhysRevD.99.035036
  [arXiv:1810.07195 [hep-ph]].
  %%CITATION = doi:10.1103/PhysRevD.99.035036;%%


%\cite{Graham:2015rva}
\bibitem{Graham:2015rva} 
  P.~W.~Graham, J.~Mardon and S.~Rajendran,
  %``Vector Dark Matter from Inflationary Fluctuations,''
  Phys.\ Rev.\ D {\bf 93}, no. 10, 103520 (2016)
  %doi:10.1103/PhysRevD.93.103520
  [arXiv:1504.02102 [hep-ph]].
  %%CITATION = doi:10.1103/PhysRevD.93.103520;%%

%\cite{Ema:2019yrd}
\bibitem{Ema:2019yrd} 
  Y.~Ema, K.~Nakayama and Y.~Tang,
  %``Production of Purely Gravitational Dark Matter: The Case of Fermion and Vector Boson,''
  arXiv:1903.10973 [hep-ph].
  %%CITATION = ARXIV:1903.10973;%%


  
  %\cite{AlonsoAlvarez:2019cgw}
\bibitem{AlonsoAlvarez:2019cgw} 
  G.~Alonso-Alvarez, T.~Hugle and J.~Jaeckel,
  %``Misalignment & Co.: (Pseudo-)scalar and vector dark matter with curvature couplings,''
  arXiv:1905.09836 [hep-ph].
  %%CITATION = ARXIV:1905.09836;%%
  
  
  
   %\cite{Dvali:2007ks}
\bibitem{Dvali:2007ks} 
  G.~Dvali, O.~Pujolas and M.~Redi,
  %``Consistent Lorentz Violation in Flat and Curved Space,''
  Phys.\ Rev.\ D {\bf 76}, 044028 (2007)
  %doi:10.1103/PhysRevD.76.044028
  [hep-th/0702117 [HEP-TH]].
  %%CITATION = doi:10.1103/PhysRevD.76.044028;%%
  
  %\cite{Himmetoglu:2008zp}
\bibitem{Himmetoglu:2008zp} 
  B.~Himmetoglu, C.~R.~Contaldi and M.~Peloso,
  %``Instability of anisotropic cosmological solutions supported by vector fields,''
  Phys.\ Rev.\ Lett.\  {\bf 102}, 111301 (2009)
  %doi:10.1103/PhysRevLett.102.111301
  [arXiv:0809.2779 [astro-ph]].
  %%CITATION = doi:10.1103/PhysRevLett.102.111301;%%
  
  %\cite{Himmetoglu:2009qi}
\bibitem{Himmetoglu:2009qi} 
  B.~Himmetoglu, C.~R.~Contaldi and M.~Peloso,
  %``Ghost instabilities of cosmological models with vector fields nonminimally coupled to the curvature,''
  Phys.\ Rev.\ D {\bf 80}, 123530 (2009)
  %doi:10.1103/PhysRevD.80.123530
  [arXiv:0909.3524 [astro-ph.CO]].
  %%CITATION = doi:10.1103/PhysRevD.80.123530;%%
  
  %\cite{Karciauskas:2010as}
\bibitem{Karciauskas:2010as} 
  M.~Karciauskas and D.~H.~Lyth,
  %``On the health of a vector field with (R A^2)/6 coupling to gravity,''
  JCAP {\bf 1011}, 023 (2010)
  %doi:10.1088/1475-7516/2010/11/023
  [arXiv:1007.1426 [astro-ph.CO]].
  %%CITATION = doi:10.1088/1475-7516/2010/11/023;%%
  

  
  
%\cite{Turner:1987bw}
\bibitem{Turner:1987bw} 
  M.~S.~Turner and L.~M.~Widrow,
  %``Inflation Produced, Large Scale Magnetic Fields,''
  Phys.\ Rev.\ D {\bf 37}, 2743 (1988).
  %doi:10.1103/PhysRevD.37.2743
  %%CITATION = doi:10.1103/PhysRevD.37.2743;%%
  
  
  

%\cite{Dimopoulos:2006ms}
\bibitem{Dimopoulos:2006ms} 
  K.~Dimopoulos,
  %``Can a vector field be responsible for the curvature perturbation in the Universe?,''
  Phys.\ Rev.\ D {\bf 74}, 083502 (2006)
  %doi:10.1103/PhysRevD.74.083502
  [hep-ph/0607229].
  %%CITATION = doi:10.1103/PhysRevD.74.083502;%%

%\cite{Dimopoulos:2008rf}
\bibitem{Dimopoulos:2008rf} 
  K.~Dimopoulos and M.~Karciauskas,
  %``Non-minimally coupled vector curvaton,''
  JHEP {\bf 0807}, 119 (2008)
  %doi:10.1088/1126-6708/2008/07/119
  [arXiv:0803.3041 [hep-th]].
  %%CITATION = doi:10.1088/1126-6708/2008/07/119;%%
  
  %\cite{Dimopoulos:2008yv}
\bibitem{Dimopoulos:2008yv} 
  K.~Dimopoulos, M.~Karciauskas, D.~H.~Lyth and Y.~Rodriguez,
  %``Statistical anisotropy of the curvature perturbation from vector field perturbations,''
  JCAP {\bf 0905}, 013 (2009)
  %doi:10.1088/1475-7516/2009/05/013
  [arXiv:0809.1055 [astro-ph]].
  %%CITATION = doi:10.1088/1475-7516/2009/05/013;%%
  
 
 
   
  
  %\cite{Dimopoulos:2007zb}
\bibitem{Dimopoulos:2007zb} 
  K.~Dimopoulos,
  %``Supergravity inspired Vector Curvaton,''
  Phys.\ Rev.\ D {\bf 76}, 063506 (2007)
  %doi:10.1103/PhysRevD.76.063506
  [arXiv:0705.3334 [hep-ph]].
  %%CITATION = doi:10.1103/PhysRevD.76.063506;%%
  
  %\cite{Dimopoulos:2009am}
\bibitem{Dimopoulos:2009am} 
  K.~Dimopoulos, M.~Karciauskas and J.~M.~Wagstaff,
  %``Vector Curvaton with varying Kinetic Function,''
  Phys.\ Rev.\ D {\bf 81}, 023522 (2010)
  %doi:10.1103/PhysRevD.81.023522
  [arXiv:0907.1838 [hep-ph]].
  %%CITATION = doi:10.1103/PhysRevD.81.023522;%%
  
  %\cite{Dimopoulos:2009vu}
\bibitem{Dimopoulos:2009vu} 
  K.~Dimopoulos, M.~Karciauskas and J.~M.~Wagstaff,
  %``Vector Curvaton without Instabilities,''
  Phys.\ Lett.\ B {\bf 683}, 298 (2010)
  %doi:10.1016/j.physletb.2009.12.024
  [arXiv:0909.0475 [hep-ph]].
  %%CITATION = doi:10.1016/j.physletb.2009.12.024;%%
  
  %\cite{Dimopoulos:2010xq}
\bibitem{Dimopoulos:2010xq} 
  J.~M.~Wagstaff and K.~Dimopoulos,
  %``Particle Production of Vector Fields: Scale Invariance is Attractive,''
  Phys.\ Rev.\ D {\bf 83}, 023523 (2011)
  %doi:10.1103/PhysRevD.83.023523
  [arXiv:1011.2517 [hep-ph]].
  %%CITATION = doi:10.1103/PhysRevD.83.023523;%%

  
  %\cite{Nelson:2011sf}
\bibitem{Nelson:2011sf} 
  A.~E.~Nelson and J.~Scholtz,
  %``Dark Light, Dark Matter and the Misalignment Mechanism,''
  Phys.\ Rev.\ D {\bf 84}, 103501 (2011)
  %doi:10.1103/PhysRevD.84.103501
  [arXiv:1105.2812 [hep-ph]].
  %%CITATION = doi:10.1103/PhysRevD.84.103501;%%
  
  
  
  %\cite{Carroll:2003st}
\bibitem{Carroll:2003st} 
  S.~M.~Carroll, M.~Hoffman and M.~Trodden,
  %``Can the dark energy equation - of - state parameter w be less than -1?,''
  Phys.\ Rev.\ D {\bf 68}, 023509 (2003)
  %doi:10.1103/PhysRevD.68.023509
  [astro-ph/0301273].
  %%CITATION = doi:10.1103/PhysRevD.68.023509;%%
  
  %\cite{Ford:1989me}
\bibitem{Ford:1989me} 
  L.~H.~Ford,
  %``Inflation Driven By A Vector Field,''
  Phys.\ Rev.\ D {\bf 40}, 967 (1989).
  %doi:10.1103/PhysRevD.40.967
  %%CITATION = doi:10.1103/PhysRevD.40.967;%%
  
  
   %\cite{Watanabe:2009ct}
\bibitem{Watanabe:2009ct} 
  M.~a.~Watanabe, S.~Kanno and J.~Soda,
  %``Inflationary Universe with Anisotropic Hair,''
  Phys.\ Rev.\ Lett.\  {\bf 102}, 191302 (2009)
  %doi:10.1103/PhysRevLett.102.191302
  [arXiv:0902.2833 [hep-th]].
  %%CITATION = doi:10.1103/PhysRevLett.102.191302;%%
    
  %\cite{Soda:2012zm}
\bibitem{Soda:2012zm} 
  J.~Soda,
  %``Statistical Anisotropy from Anisotropic Inflation,''
  Class.\ Quant.\ Grav.\  {\bf 29}, 083001 (2012)
  %doi:10.1088/0264-9381/29/8/083001
  [arXiv:1201.6434 [hep-th]].
  %%CITATION = doi:10.1088/0264-9381/29/8/083001;%%
  
    %\cite{Maleknejad:2012fw}
\bibitem{Maleknejad:2012fw} 
  A.~Maleknejad, M.~M.~Sheikh-Jabbari and J.~Soda,
  %``Gauge Fields and Inflation,''
  Phys.\ Rept.\  {\bf 528}, 161 (2013)
  %doi:10.1016/j.physrep.2013.03.003
  [arXiv:1212.2921 [hep-th]].
  %%CITATION = doi:10.1016/j.physrep.2013.03.003;%%

  %\cite{Liddle:2000cg}
\bibitem{Liddle:2000cg} 
  A.~R.~Liddle and D.~H.~Lyth,
  ``Cosmological inflation and large scale structure,''
  Cambridge, UK: Univ. Pr. (2000) 400 p
  
  
  
  %\cite{Ratra:1991bn}
\bibitem{Ratra:1991bn} 
  B.~Ratra,
  %``Cosmological 'seed' magnetic field from inflation,''
  Astrophys.\ J.\  {\bf 391}, L1 (1992).
  %doi:10.1086/186384
  %%CITATION = doi:10.1086/186384;%%
  
  %\cite{Giovannini:2001nh}
\bibitem{Giovannini:2001nh} 
  M.~Giovannini,
  %``On the variation of the gauge couplings during inflation,''
  Phys.\ Rev.\ D {\bf 64}, 061301 (2001)
  %doi:10.1103/PhysRevD.64.061301
  [astro-ph/0104290].
  %%CITATION = doi:10.1103/PhysRevD.64.061301;%%
  
  %\cite{Bamba:2003av}
\bibitem{Bamba:2003av} 
  K.~Bamba and J.~Yokoyama,
  %``Large scale magnetic fields from inflation in dilaton electromagnetism,''
  Phys.\ Rev.\ D {\bf 69}, 043507 (2004)
  %doi:10.1103/PhysRevD.69.043507
  [astro-ph/0310824].
  %%CITATION = doi:10.1103/PhysRevD.69.043507;%%
  
  %\cite{Kandus:2010nw}
\bibitem{Kandus:2010nw} 
  A.~Kandus, K.~E.~Kunze and C.~G.~Tsagas,
  %``Primordial magnetogenesis,''
  Phys.\ Rept.\  {\bf 505}, 1 (2011)
  %doi:10.1016/j.physrep.2011.03.001
  [arXiv:1007.3891 [astro-ph.CO]].
  %%CITATION = doi:10.1016/j.physrep.2011.03.001;%%
  
  %\cite{Demozzi:2009fu}
\bibitem{Demozzi:2009fu} 
  V.~Demozzi, V.~Mukhanov and H.~Rubinstein,
  %``Magnetic fields from inflation?,''
  JCAP {\bf 0908}, 025 (2009)
  %doi:10.1088/1475-7516/2009/08/025
  [arXiv:0907.1030 [astro-ph.CO]].
  %%CITATION = doi:10.1088/1475-7516/2009/08/025;%%
  
  %\cite{Kanno:2009ei}
\bibitem{Kanno:2009ei} 
  S.~Kanno, J.~Soda and M.~a.~Watanabe,
  %``Cosmological Magnetic Fields from Inflation and Backreaction,''
  JCAP {\bf 0912}, 009 (2009)
  %doi:10.1088/1475-7516/2009/12/009
  [arXiv:0908.3509 [astro-ph.CO]].
  %%CITATION = doi:10.1088/1475-7516/2009/12/009;%%
  
  %\cite{Fujita:2012rb}
\bibitem{Fujita:2012rb} 
  T.~Fujita and S.~Mukohyama,
  %``Universal upper limit on inflation energy scale from cosmic magnetic field,''
  JCAP {\bf 1210}, 034 (2012)
  %doi:10.1088/1475-7516/2012/10/034
  [arXiv:1205.5031 [astro-ph.CO]].
  %%CITATION = doi:10.1088/1475-7516/2012/10/034;%%
  
  %\cite{Akrami:2018odb}
\bibitem{Akrami:2018odb} 
  Y.~Akrami {\it et al.} [Planck Collaboration],
  %``Planck 2018 results. X. Constraints on inflation,''
  arXiv:1807.06211 [astro-ph.CO].
  %%CITATION = ARXIV:1807.06211;%%
  
    %\cite{Namba:2012gg}
\bibitem{Namba:2012gg} 
  R.~Namba,
  %``Curvature Perturbations from a Massive Vector Curvaton,''
  Phys.\ Rev.\ D {\bf 86}, 083518 (2012)
  %doi:10.1103/PhysRevD.86.083518
  [arXiv:1207.5547 [astro-ph.CO]].
  %%CITATION = doi:10.1103/PhysRevD.86.083518;%%  
  
  %\cite{Destri:2007pv}
\bibitem{Destri:2007pv} 
  C.~Destri, H.~J.~de Vega and N.~G.~Sanchez,
  %``MCMC analysis of WMAP3 and SDSS data points to broken symmetry inflaton potentials and provides a lower bound on the tensor to scalar ratio,''
  Phys.\ Rev.\ D {\bf 77}, 043509 (2008)
  %doi:10.1103/PhysRevD.77.043509
  [astro-ph/0703417].
  %%CITATION = doi:10.1103/PhysRevD.77.043509;%%
  
  %\cite{Nakayama:2013jka}
\bibitem{Nakayama:2013jka} 
  K.~Nakayama, F.~Takahashi and T.~T.~Yanagida,
  %``Polynomial Chaotic Inflation in the Planck Era,''
  Phys.\ Lett.\ B {\bf 725}, 111 (2013)
  %doi:10.1016/j.physletb.2013.06.050
  [arXiv:1303.7315 [hep-ph]].
  %%CITATION = doi:10.1016/j.physletb.2013.06.050;%%
  
  %\cite{Nakayama:2013txa}
\bibitem{Nakayama:2013txa} 
  K.~Nakayama, F.~Takahashi and T.~T.~Yanagida,
  %``Polynomial Chaotic Inflation in Supergravity,''
  JCAP {\bf 1308}, 038 (2013)
  %doi:10.1088/1475-7516/2013/08/038
  [arXiv:1305.5099 [hep-ph]].
  %%CITATION = doi:10.1088/1475-7516/2013/08/038;%%
  
  
  
  
   %\cite{Sanchez:2013zst}
\bibitem{Sanchez:2013zst} 
  J.~C.~Bueno Sanchez and K.~Dimopoulos,
  %``Inflationary buildup of a vector field condensate and its cosmological consequences,''
  JCAP {\bf 1401}, 012 (2014)
  %doi:10.1088/1475-7516/2014/01/012
  [arXiv:1308.3739 [hep-ph]].
  %%CITATION = doi:10.1088/1475-7516/2014/01/012;%%
  
  
  
  %\cite{Lyth:2004gb}
\bibitem{Lyth:2004gb} 
  D.~H.~Lyth, K.~A.~Malik and M.~Sasaki,
  %``A General proof of the conservation of the curvature perturbation,''
  JCAP {\bf 0505}, 004 (2005)
  %doi:10.1088/1475-7516/2005/05/004
  [astro-ph/0411220].
  %%CITATION = doi:10.1088/1475-7516/2005/05/004;%%
  
  %\cite{Wands:2000dp}
\bibitem{Wands:2000dp} 
  D.~Wands, K.~A.~Malik, D.~H.~Lyth and A.~R.~Liddle,
  %``A New approach to the evolution of cosmological perturbations on large scales,''
  Phys.\ Rev.\ D {\bf 62}, 043527 (2000)
  %doi:10.1103/PhysRevD.62.043527
  [astro-ph/0003278].
  %%CITATION = doi:10.1103/PhysRevD.62.043527;%%
  
  %\cite{Kawasaki:2008sn}
\bibitem{Kawasaki:2008sn} 
  M.~Kawasaki, K.~Nakayama, T.~Sekiguchi, T.~Suyama and F.~Takahashi,
  %``Non-Gaussianity from isocurvature perturbations,''
  JCAP {\bf 0811}, 019 (2008)
  %doi:10.1088/1475-7516/2008/11/019
  [arXiv:0808.0009 [astro-ph]].
  %%CITATION = doi:10.1088/1475-7516/2008/11/019;%%
  
  %\cite{Langlois:2008vk}
\bibitem{Langlois:2008vk} 
  D.~Langlois, F.~Vernizzi and D.~Wands,
  %``Non-linear isocurvature perturbations and non-Gaussianities,''
  JCAP {\bf 0812}, 004 (2008)
  %doi:10.1088/1475-7516/2008/12/004
  [arXiv:0809.4646 [astro-ph]].
  %%CITATION = doi:10.1088/1475-7516/2008/12/004;%%
  
  %\cite{Abazajian:2016yjj}
\bibitem{Abazajian:2016yjj} 
  K.~N.~Abazajian {\it et al.} [CMB-S4 Collaboration],
  %``CMB-S4 Science Book, First Edition,''
  arXiv:1610.02743 [astro-ph.CO].
  %%CITATION = ARXIV:1610.02743;%%
  
%\cite{Arvanitaki:2009fg}
\bibitem{Arvanitaki:2009fg} 
  A.~Arvanitaki, S.~Dimopoulos, S.~Dubovsky, N.~Kaloper and J.~March-Russell,
  %``String Axiverse,''
  Phys.\ Rev.\ D {\bf 81}, 123530 (2010)
  %doi:10.1103/PhysRevD.81.123530
  [arXiv:0905.4720 [hep-th]].
  %%CITATION = doi:10.1103/PhysRevD.81.123530;%%
  
  %\cite{Arvanitaki:2010sy}
\bibitem{Arvanitaki:2010sy} 
  A.~Arvanitaki and S.~Dubovsky,
  %``Exploring the String Axiverse with Precision Black Hole Physics,''
  Phys.\ Rev.\ D {\bf 83}, 044026 (2011)
  %doi:10.1103/PhysRevD.83.044026
  [arXiv:1004.3558 [hep-th]].
  %%CITATION = doi:10.1103/PhysRevD.83.044026;%%
  
  %\cite{Pani:2012vp}
\bibitem{Pani:2012vp} 
  P.~Pani, V.~Cardoso, L.~Gualtieri, E.~Berti and A.~Ishibashi,
  %``Black hole bombs and photon mass bounds,''
  Phys.\ Rev.\ Lett.\  {\bf 109}, 131102 (2012)
  %doi:10.1103/PhysRevLett.109.131102
  [arXiv:1209.0465 [gr-qc]].
  %%CITATION = doi:10.1103/PhysRevLett.109.131102;%%
  
  %\cite{Brito:2015oca}
\bibitem{Brito:2015oca} 
  R.~Brito, V.~Cardoso and P.~Pani,
  %``Superradiance : Energy Extraction, Black-Hole Bombs and Implications for Astrophysics and Particle Physics,''
  Lect.\ Notes Phys.\  {\bf 906}, pp.1 (2015)
  %doi:10.1007/978-3-319-19000-6
  [arXiv:1501.06570 [gr-qc]].
  %%CITATION = doi:10.1007/978-3-319-19000-6;%%
  
  %\cite{Horns:2012jf}
\bibitem{Horns:2012jf} 
  D.~Horns, J.~Jaeckel, A.~Lindner, A.~Lobanov, J.~Redondo and A.~Ringwald,
  %``Searching for WISPy Cold Dark Matter with a Dish Antenna,''
  JCAP {\bf 1304}, 016 (2013)
  %doi:10.1088/1475-7516/2013/04/016
  [arXiv:1212.2970 [hep-ph]].
  %%CITATION = doi:10.1088/1475-7516/2013/04/016;%%
  
  %\cite{Parker:2013fxa}
\bibitem{Parker:2013fxa} 
  S.~R.~Parker, J.~G.~Hartnett, R.~G.~Povey and M.~E.~Tobar,
  %``Cryogenic resonant microwave cavity searches for hidden sector photons,''
  Phys.\ Rev.\ D {\bf 88}, 112004 (2013)
  %doi:10.1103/PhysRevD.88.112004
  [arXiv:1410.5244 [hep-ex]].
  %%CITATION = doi:10.1103/PhysRevD.88.112004;%%
  
  %\cite{Chaudhuri:2014dla}
\bibitem{Chaudhuri:2014dla} 
  S.~Chaudhuri, P.~W.~Graham, K.~Irwin, J.~Mardon, S.~Rajendran and Y.~Zhao,
  %``Radio for hidden-photon dark matter detection,''
  Phys.\ Rev.\ D {\bf 92}, no. 7, 075012 (2015)
  %doi:10.1103/PhysRevD.92.075012
  [arXiv:1411.7382 [hep-ph]].
  %%CITATION = doi:10.1103/PhysRevD.92.075012;%%
  
  %\cite{Hochberg:2016ajh}
\bibitem{Hochberg:2016ajh} 
  Y.~Hochberg, T.~Lin and K.~M.~Zurek,
  %``Detecting Ultralight Bosonic Dark Matter via Absorption in Superconductors,''
  Phys.\ Rev.\ D {\bf 94}, no. 1, 015019 (2016)
  %doi:10.1103/PhysRevD.94.015019
  [arXiv:1604.06800 [hep-ph]].
  %%CITATION = doi:10.1103/PhysRevD.94.015019;%%
  
  %\cite{Hochberg:2016sqx}
\bibitem{Hochberg:2016sqx} 
  Y.~Hochberg, T.~Lin and K.~M.~Zurek,
  %``Absorption of light dark matter in semiconductors,''
  Phys.\ Rev.\ D {\bf 95}, no. 2, 023013 (2017)
  %doi:10.1103/PhysRevD.95.023013
  [arXiv:1608.01994 [hep-ph]].
  %%CITATION = doi:10.1103/PhysRevD.95.023013;%%
    
  %\cite{Arvanitaki:2017nhi}
\bibitem{Arvanitaki:2017nhi} 
  A.~Arvanitaki, S.~Dimopoulos and K.~Van Tilburg,
  %``Resonant absorption of bosonic dark matter in molecules,''
  Phys.\ Rev.\ X {\bf 8}, no. 4, 041001 (2018)
  %doi:10.1103/PhysRevX.8.041001
  [arXiv:1709.05354 [hep-ph]].
  %%CITATION = doi:10.1103/PhysRevX.8.041001;%%
  
  %\cite{Baryakhtar:2018doz}
\bibitem{Baryakhtar:2018doz} 
  M.~Baryakhtar, J.~Huang and R.~Lasenby,
  %``Axion and hidden photon dark matter detection with multilayer optical haloscopes,''
  Phys.\ Rev.\ D {\bf 98}, no. 3, 035006 (2018)
  %doi:10.1103/PhysRevD.98.035006
  [arXiv:1803.11455 [hep-ph]].
  %%CITATION = doi:10.1103/PhysRevD.98.035006;%%
  
  
  
  
  
   %\cite{Maleknejad:2011jw}
\bibitem{Maleknejad:2011jw} 
  A.~Maleknejad and M.~M.~Sheikh-Jabbari,
  %``Gauge-flation: Inflation From Non-Abelian Gauge Fields,''
  Phys.\ Lett.\ B {\bf 723}, 224 (2013)
  %doi:10.1016/j.physletb.2013.05.001
  [arXiv:1102.1513 [hep-ph]].
  %%CITATION = doi:10.1016/j.physletb.2013.05.001;%%
  
  %\cite{Maleknejad:2011sq}
\bibitem{Maleknejad:2011sq} 
  A.~Maleknejad and M.~M.~Sheikh-Jabbari,
  %``Non-Abelian Gauge Field Inflation,''
  Phys.\ Rev.\ D {\bf 84}, 043515 (2011)
  %doi:10.1103/PhysRevD.84.043515
  [arXiv:1102.1932 [hep-ph]].
  %%CITATION = doi:10.1103/PhysRevD.84.043515;%%
  
 
  
  %\cite{Nakayama:2020rka}
\bibitem{Nakayama:2020rka}
K.~Nakayama,
%``Constraint on Vector Coherent Oscillation Dark Matter with Kinetic Function,''
JCAP \textbf{08}, 033 (2020)
%doi:10.1088/1475-7516/2020/08/033
[arXiv:2004.10036 [hep-ph]].
  
  
  
  

%%%%%%%%%%%%%%%%%%%%%%%%%%%%%%%%%%%%%%%%%%%%%%%%%%
\end{thebibliography}
\end{document}